\documentclass[twocolumn]{aastex62}
\usepackage{natbib}
\usepackage{gensymb}
\usepackage{enumitem}
\usepackage{graphicx}
\usepackage{multirow}
\usepackage[normalem]{ulem}
\usepackage{soul}
\usepackage{tabularx,booktabs}
\usepackage[caption=false]{subfig}
\begin{document} 

\title{\Large{Supervised convolutional neural networks for classification of flaring and nonflaring active regions using line-of-sight magnetograms}}
\correspondingauthor{Dattaraj Dhuri}
\email{dattaraj.dhuri@tifr.res.in}

\author{Shamik Bhattacharjee}
\affiliation{Department of Astronomy and Astrophysics, Tata Institute of Fundamental Research, Mumbai, India 400005}

\author{Rasha Alshehhi}
\affiliation{Center of Space Science, New York University Abu Dhabi, Abu Dhabi, United Arab Emirates}

\author{Dattaraj B. Dhuri}
\affiliation{Department of Astronomy and Astrophysics, Tata Institute of Fundamental Research, Mumbai, India 400005}

\author{Shravan M. Hanasoge}
\affiliation{Department of Astronomy and Astrophysics, Tata Institute of Fundamental Research, Mumbai, India 400005}
\affiliation{Center of Space Science, New York University Abu Dhabi, Abu Dhabi, United Arab Emirates}

\begin{abstract}
Solar flares are explosions in the solar atmosphere that release intense bursts of short-wavelength radiation and are capable of producing severe space-weather consequences. Flares release free energy built up in coronal fields, which are rooted in active regions (ARs) on the photosphere, via magnetic reconnection. The exact processes that lead to reconnection are not fully known and therefore reliable forecasting of flares is challenging. Recently, photospheric magnetic-field data has been extensively analysed using machine learning (ML) and these studies suggest that flare-forecasting accuracy does not strongly depend on how long in advance flares are predicted \citep{bobraflareprediction,Raboonik2016,Huang2018}. Here, we use ML to understand the evolution of AR magnetic fields before and after flares. We explicitly train convolutional neural networks (CNNs) to classify SDO/HMI line-of-sight magnetograms into ARs producing at least one M- or X-class flare or as nonflaring. We find that flaring ARs remain in flare-productive states --- marked by \textit{recall} $>60\%$ with a peak of $\sim 80\%$ --- days before and after flares. We use occlusion maps and statistical analysis to show that the CNN pays attention to regions between the opposite polarities from ARs and the CNN output is dominantly decided by the total unsigned line-of-sight flux of ARs. Using synthetic bipole magnetograms, we find spurious dependencies of the CNN output on magnetogram dimensions for a given bipole size. Our results suggest that it is important to use CNN designs that eliminate such artifacts in CNN applications for processing magnetograms and, in general, solar image data.
\end{abstract}

\section{Introduction}
Solar flares release free energy built up in the coronal magnetic fields in the form of intense short-wavelength radiation. Flare intensity is measured in terms of X-ray flux and major flares, i.e. M- and X-class flares, produce peak X-ray flux of $>10^{-5}\ \textrm{W-m}^{-2}$ and $>10^{-4}\ \textrm{W-m}^{-2}$ respectively. The short-wavelength radiation released in flares causes disruptions in GPS communication, radio blackouts and poses health hazards to astronauts and flight crew. Reliable forecasting of flares and other space-weather events is, therefore, necessary \citep{Eastwood2017}.

Coronal magnetic fields are energized by the emergence of magnetic flux from the solar interior and subsequent build up of electric current \citep{Cheung2014,Stein2012,Leka1996}. Flares occur as a consequence of magnetic reconnection of coronal fields \citep{Shibata2011,Su2013}. Over the past few decades, several case studies and statistical studies have focused on the analysis of photospheric magnetic-field, obtained from space as well as ground-based observatories, to understand flare precursors for reliable forecasting \citep{SCHRIJVER2009739,Leka2008,Wang2015}. Features such as continuously emerging flux \citep{Nitta2001}, strong polarity inversion line \citep{Schrijver2007} and accumulation of electric current and magnetic helicity \cite{Park2008,Kontogiannis2017} are found to be strongly correlated with flaring activity. However, no single measure of photospheric magnetic field is sufficient for reliably forecasting flares \citep{Leka2008}. Operational flare forecasts rely therefore on the analysis of AR magnetograms and coronal images by human experts \citep{McIntosh1990,Rust1994,crown2012validation} and reliable automated forecasting of flares is yet to be achieved \citep{AllClear}.

\textit{Helioseismic and Magnetic Imager} (HMI) \citep{Scherrer2012} onboard NASA's \textit{Solar Dynamics Observatory} (SDO) \citep{Pesnell-etall2012} provides high-resolution photospheric vector-magnetic-field images. With the availability of machine learning (ML) techniques \citep{hastie01statisticallearning}, these data have been extensively analyzed for improving flare forecasting. ML approaches have primarily relied on using magnetic-field features calculated from vector-magnetograms, such as space-weather HMI active region patches (SHARPS) \citep{Bobra2014}, known to be correlated with flare activity. These magnetic-field features describe average characteristics of ARs and are analysed by a variety of ML algorithms trained for forecasting flares \citep{Ahmed2013,bobraflareprediction,Raboonik2016,Nishizuka2017,Jonas2018}. Overall, these forecasts have yielded statistically superior performance than those based on subjective analyses of ARs \citep{crown2012validation}. The leading contributors for flare forecasting in these ML studies have been AR magnetic-field features corresponding to extensive AR properties, e.g. total unsigned magnetic flux \citep{bobraflareprediction,Dhuri2019}.

Rather than only considering AR-averaged magnetic-field features, advanced ML techniques such as convolutional neural networks (CNNs) \citep{Goodfellow2016,Krizhevsky2012,LeCun2015} provide an opportunity to directly process AR magnetograms and characterise AR morphological features correlated with flares. CNNs trained on magnetograms may automatically extract subtle and localized features in AR magnetic fields that are precursors to flares, thereby improving flare forecasts and our understanding of flare mechanisms. For instance, \cite{Huang2018} used line-of-sight magnetograms to train CNNs for forecasting M- and X-class flares. Their result suggests that forecasting accuracy does not reduce appreciably as the forward-looking-time, i.e., time in advance of the flare, is increased. This is consistent with earlier studies using features derived from AR magnetograms \citep{bobraflareprediction,Raboonik2016}.

\begin{table}[b]
\begin{tabular}{lcc}
&&\\
\hline
 & May'10 - Sep'15 & Oct'15 - Aug'18 \\
                   & Train \& Val & Test \\
\hline
\# flaring ARs               &    161        &  20      \\
\# non-flaring ARs           &  696              &   191 \\
\# M- \& X-class flares            &   627           &  106 \\
\hline
\end{tabular}
\caption{Active region (AR) dataset used for classification of flaring and nonflaring ARs using convolutional neural networks (CNNs).}
\label{tab:dataset}
\end{table}
\cite{Dhuri2019} explicitly trained support vector machines (SVMs) to classify SHARP features derived from flaring and nonflaring ARs. They found that flaring ARs remain in flare-productive states days before and after M- and X-class flares, marked by distinctly high values of extensive AR features. In the present work, we use supervised learning to train CNNs to distinguish between  line-of-sight magnetograms of flaring and nonflaring ARs. The CNN builds a correlation between spatial patterns identified in AR magnetograms and flaring activity. Following \cite{Dhuri2019}, we explicitly study how machine correlation changes days before and after flares. Notwithstanding their success in performing classification and pattern detection tasks, it is challenging to understand the operation and components of CNNs and deep neural networks. Here, we use statistical analysis of the machine correlation as well as occlusion maps to infer morphological patterns detected by the CNN and interpret machine performance. Using synthetic magnetograms, we find that the CNN output depends on systematic factors arising as a consequence of unequal sizes of AR magnetograms. 

This paper is organized as follows. In section \ref{sec:data}, we detail line-of-sight magnetic-field data used for the analysis. In section \ref{sec:method}, we explain the CNN architectures used  --- a simple CNN with two convolutional layers as a baseline model and another with inception modules similar to GoogleNet \citep{GoogleNet2015} that incorporates different spatial convolution filters in a single convolutional layer. In section \ref{sec:results}, we compare the performances of the two CNN architectures for the classification of flaring and nonflaring ARs. We explain in detail statistical analyses of CNN outputs that are performed to understand the CNN operation. We also compare the results of the CNN with the classification results of \cite{Dhuri2019} using vector-magnetic-field features. Using synthetic data, we trace systematic errors in the CNN classification to unequal AR sizes. We present occlusion maps obtained to highlight the morphological patterns learned by the CNN. In section \ref{sec:summary}, we summarise our findings.

\section{Data \label{sec:data}}
\begin{figure*}[t]
\centering
\includegraphics[width=0.75\textwidth]{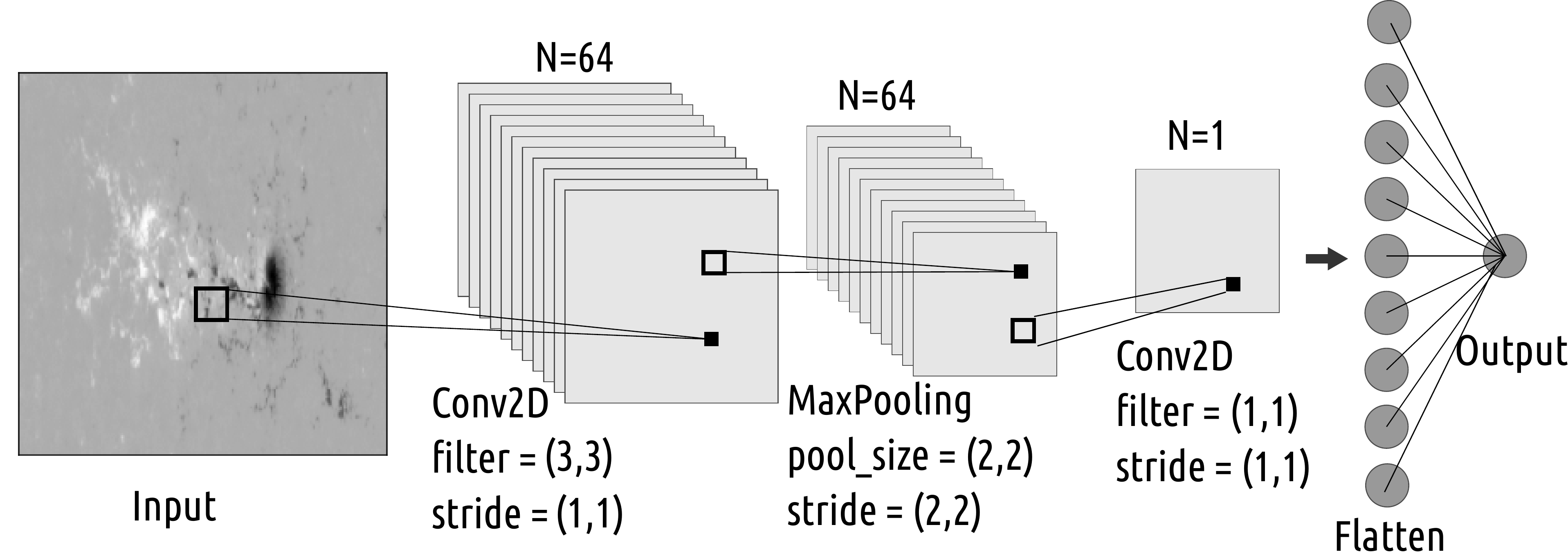}
\caption{A simple convolutional neural network (CNN), referred to as the CNN-1, used for the classification of line-of-sight magnetograms of flaring and nonflaring active regions (ARs). The CNN-1 consists of two Conv2D layers as shown, which are made up of 2D convolutional filters that scan over the image for magnetic field features. Each convolutional layer is followed by a max-pooling layer, which downsamples the image. The output of the second convolutional layer is {\it flattened} and processed by a fully connected (FC) layer of neurons. The FC layer is connected to the output neuron. The CNN-1 serves as a baseline model for the classification of flaring and nonflaring ARs.}
\label{fig:cnn1}
\end{figure*}

We use line-of-sight magnetograms provided by the {\it Solar Dynamics Observatory} (SDO) and a solar flare events catalogue provided by the {\it Geostationary Operational Environmental Satellite} (GOES). These datasets are publicly available.

Since 2010, SDO monitors solar activity by imaging the solar surface and atmosphere. {\it Helioseismic and Magnetic Imager} (HMI) onboard SDO yields full-disk vector-magnetograms every 12 minutes with a plate scale of $0.5\,{\rm arcsecs}$ ($\sim 380\,{\rm km}$ at the disk center). From full-disk magnetograms, AR patches are automatically detected and tracked as they rotate across the visible solar disk. These AR patches are available among HMI-derived data products {\it Space-weather HMI Active Region Patches} (SHARPs) \citep{Bobra2014}. To eliminate projection effects, magnetograms are remapped on a cylindrical equal-area (CEA) grid. The CEA magnetograms for each AR in SHARPs data series are available at a cadence of 12 minutes. GOES provides a catalogue of solar flares (since 1986) and also identifies ARs that produce them according to the National Oceanic and Atmospheric Administrations' (NOAA) AR numbering scheme. ARs identified by SHARPs may contain more than one AR as per the NOAA definition \citep{Bobra2014}. We consider an AR, as identified in the SHARP data series, as flaring if it contains any of the NOAA ARs that produce at least one M- or X-class flare during their passage across the visible solar disk. Otherwise, ARs are classified as nonflaring. For every AR, we consider the magnetogram samples taken at every 1 hour.

The QUALITY keyword in the SHARP dataset indicates observations and measurement conditions for the magnetograms \citep{Bobra2014}. We consider only those measurements for which QUALITY $\leq$ 10000 in hexadecimal, indicating that Stokes vectors are reliable. The HMI instrument-noise level is sensitive to the relative velocity between the SDO and Sun. Therefore, we only consider observations obtained when the relative velocity $< 3500\,{\rm m/s}$ \citep{Hoeksema2014} and within $\pm 45\degree$ of the central meridian. We only include ARs from the SHARPs series with maximum area $>25\,{\rm Mm}^2$. This eliminates nonflaring ARs with very small sizes and does not affect the flaring AR population. We use the SHARP data series from the publicly accessible JSOC data server at \href{http://jsoc.stanford.edu/}{http://jsoc.stanford.edu/} and the GOES flare catalogue via python solar physics library Sunpy \citep{sunpy}.

We consider flaring and nonflaring ARs between May 2010 and Aug 2018. We chronologically split the available data into two parts: ARs between May 2010 - Sep 2015 are used for training and validation of CNNs and the remaining ARs, between Oct 2015 - Apr 2018, are used as test data. The number of ARs considered in the study is listed in Table \ref{tab:dataset}. We train CNNs to classify ARs as flaring, labelled 1, and nonflaring, labelled 0. Since flaring activity depends on the solar cycle variation, chronologically splitting the data for training and test may introduce a bias. Indeed, the ratio of the number of flaring to nonflaring ARs in the test set is approximately half the training-validation set. Thus, for the test data, identification of flaring ARs is expected to be more challenging and identification of nonflaring ARs is expected to be easier for trained CNNs.

\begin{figure}[t]
\centering
\includegraphics[width=0.30\textwidth]{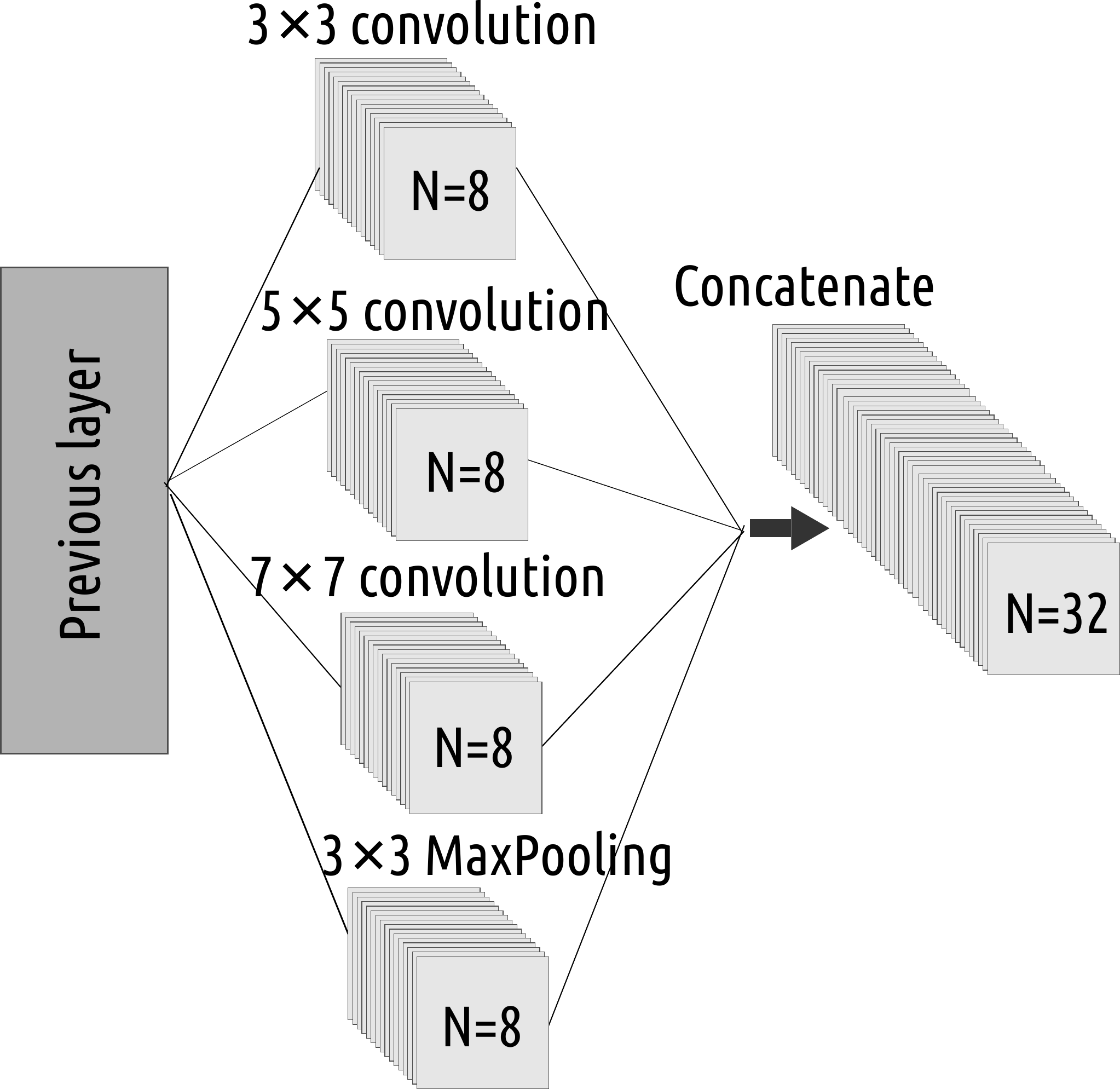}
\caption{The inception module used in the CNN-2 (Figure \ref{fig:cnn2}). The inception module comprises of 2D convolutional filters of three sizes  --- $3 \times 3$, $5 \times 5$ and $7 \times 7$. It also comprises a $3 \times 3$ max-pooling layer. The convolution filters of different sizes are sensitive to magnetic-field features of different length-scales. Outputs of the three convolutional layers and the max-pooling layer are concatenated to be fed as the input to the next layer.}
\label{fig:cnn2I}
\end{figure}

\section{Method \label{sec:method}}
\begin{figure*}[t]
\centering
\includegraphics[width=1\textwidth]{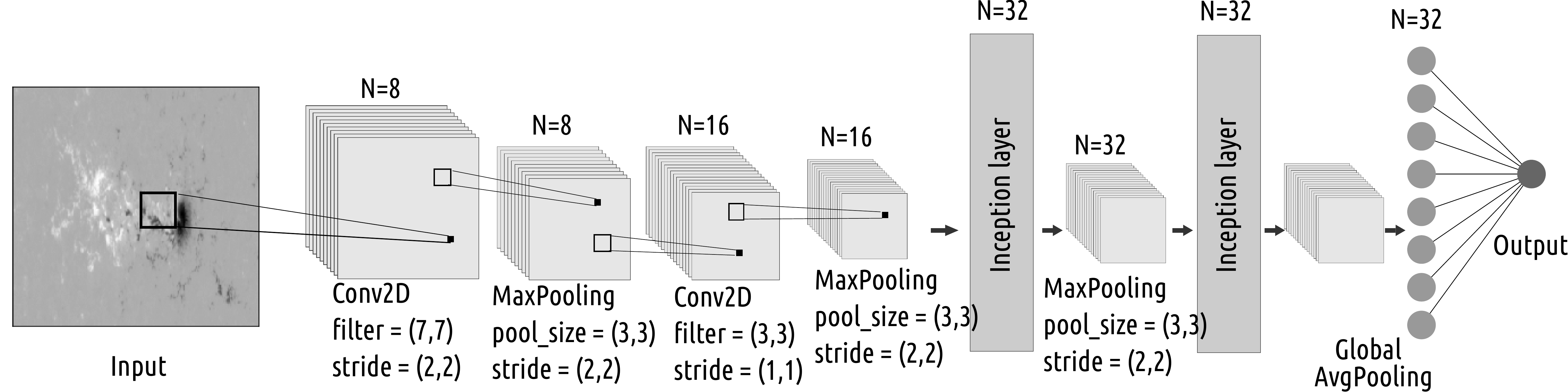}
\caption{A complex convolutional neural network (CNN), referred to as the CNN-2, used for the classification of line-of-sight magnetograms of flaring and nonflaring active regions (ARs). The CNN-2 comprises two layers of 2D convolutional filters of sizes $7 \times 7$ and $3 \times 3$ respectively, followed by two inception modules (Figure \ref{fig:cnn2I}). The two convolutional layers and the first inception module are followed by a $3 \times 3$ max-pooling layer each for downsampling. The final inception module is followed by a global-average-pooling layer that outputs the average value from the input. The output of the global-average-pooling layer is connected to the final-output neuron.}
\label{fig:cnn2}
\end{figure*}

We use CNNs (Convolutional Neural Networks) to distinguish between line-of-sight magnetograms of flaring and nonflaring ARs. CNNs, in contrast to widely used fully connected networks \citep{hastie01statisticallearning}, use convolutional filters (also known as kernels) to scan   input images and detect patterns. The convolutional filters that slide over the image are in the form of ${\rm M \times M}$ neurons where M is very small compared to the input-image dimensions (typically a $3 \times 3$ filter e.g. in \cite{Simonyan2014}).  Hence, there are far fewer numbers of parameters as compared to a fully connected network. CNNs have been hugely successful in finding patterns in images for performing tasks such as image classification, object detection, etc. \citep{LeCun2015,Goodfellow2016}.

A 2D convolutional filter of ${\rm M \times M}$ neurons sequentially scans over input images, at a time processing a ${\rm M \times M}$-pixel sub-region as per the neuron activation function (see Appendix \ref{app:App1}). Each neuron in the CNN filter yields $y=f\left(\sum_i{w_i x_i + b}\right)$,  where $\textbf{x}$ are  inputs  with  weights $\textbf{w}$, $b$ is the filter bias, and $f$ is the activation function. CNNs also comprise max-pooling layers.  A max-pooling filter of size ${\rm N \times N}$-pixel downsamples the input by factor N, picking out the maximum value of the ${\rm N \times N}$-pixel sub-region. As a result, each convolutional layer is sensitive to features of larger length scales (by factor N) in comparison to the preceding layer.

The CNN architecture for a given problem can vary from simple to deep and complex in terms of number and design of convolutional layers. Here, we train two types of CNNs --- a simple architecture that serves as a baseline model and a complex architecture using inception modules similar to the one used in the GoogleNet \citep{GoogleNet2015}. The details of the CNN architectures are as follows. 

\begin{itemize}
    \item {\bf CNN-1:} This comprises two convolutional layers, followed by one fully connected layer and the output layer. The first convolution layer is followed by a 2 $\times$ 2 max-pooling layer. A schematic illustration of the network architecture is shown in Figure \ref{fig:cnn1}. 
    
    \item {\bf CNN-2:} This is a complex architecture using inception modules similar to inception V1 modules from the GoogleNet \citep{GoogleNet2015}. Typically, in a convolution layer, we use filters of fixed size e.g. $3 \times 3$ \citep{Simonyan2014}. Depending on the problem, a particular filter size may work the best. However, the input images may contain features correlated with flaring activity over a variety of length-scales.  The inception V1 modules are designed for a situation like this. The inception module comprises convolution filters of different sizes. The output of these convolution operations is concatenated and fed as an input to the next layer. We use the inception module with three different convolution filters and one max-pooling filter as shown in Figure \ref{fig:cnn2I}. The complete CNN-2 architecture is shown in Figure \ref{fig:cnn2}. It consists of two conventional convolutional layers followed by two inception modules. The two conventional convolutional layers and the first inception module is followed by a max-pooling layer. The final inception module is followed by a global-average-pooling layer which is then connected to the output neuron.
\end{itemize}

We use minibatch stochastic-gradient descent \citep{hastie01statisticallearning} to train the CNNs. The CNNs process the input magnetograms and output a number between 0 and 1 which may be interpreted as a probability of the magnetogram belonging to the flaring population. The output is compared with the true label 0 and 1 for nonflaring and flaring ARs respectively and a measure of misfit, i.e. loss, is calculated (see Appendix \ref{app:App1}). During the training, the weights and biases of CNNs are tuned to minimize the loss using gradient descent. For effective training, {\it hyper-parameters} such as the learning rate $lr$ and minibatch size $n_{\rm BS}$ also need to be tuned. We search for and fix the learning rate and minibatch size such that the CNN classification performance is maximized. The CNN output, a number between 0 and 1, is thresholded at 0.5 to obtain the predicted label 0 and 1 for nonflaring and flaring respectively. This CNN output is categorized as follows.
\begin{itemize}
\item True Positives (TPs) - Sub-population of flaring magnetograms classified as flaring (1).
\item True Negatives (TNs) - Sub-population of nonflaring magnetograms classified as nonflaring (0).
\item False Positives (FPs) - Sub-population of nonflaring magnetograms classified as flaring (1).
\item False Negatives (FNs) - Sub-population of flaring magnetograms classified as nonflaring (0).
\end{itemize}
Since the number of nonflaring ARs is approximately 5 times larger than the number of flaring ARs, the classification problem considered here is class imbalanced. Therefore, we use performance measures that reliably capture the classification performance of the minority class i.e. the positive class. \citep{bobraflareprediction}. {\it Recall} measures the fraction of accurately classified samples for a particular class. For the positive class (flaring ARs), {\it recall} = $TP/\left(TP + FN\right)$. A CNN optimized for yielding high {\it recall} may exhibit a tendency to classify samples as positive. A better measure therefore is {\it True Skill Statistics} (TSS). {\it TSS} is calculated by subtracting the {\it false positive rate} from {\it recall} i.e. $TSS = TP/\left(TP + FN\right) - FP/\left(TN+FP\right)$. The value of {\it TSS} is 1 for the perfect classification and 0 for completely random classification. We use {\it recall} and {\it TSS} to measure the CNN classification performance.

\section{Results and Discussion \label{sec:results}}
\subsection{Training}
\begin{figure}[b]
\centering
\subfloat{
  \includegraphics[trim = {0.7cm 0.6cm 0.5cm 0.5cm},clip,width=0.325\textwidth]{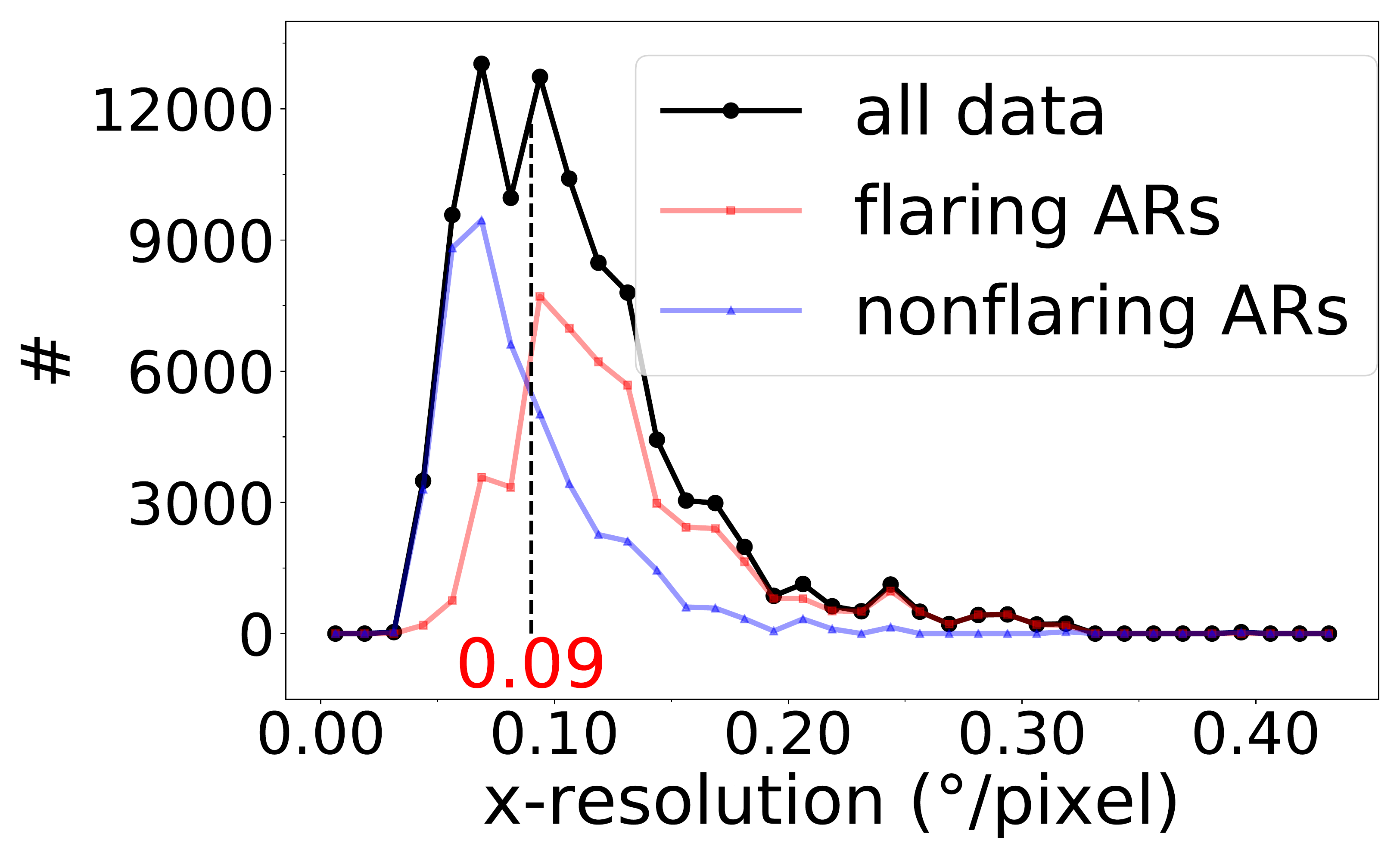}
}\\
\subfloat{
  \includegraphics[trim = {0.7cm 0.6cm 0.5cm 0.5cm},clip,width=0.325\textwidth]{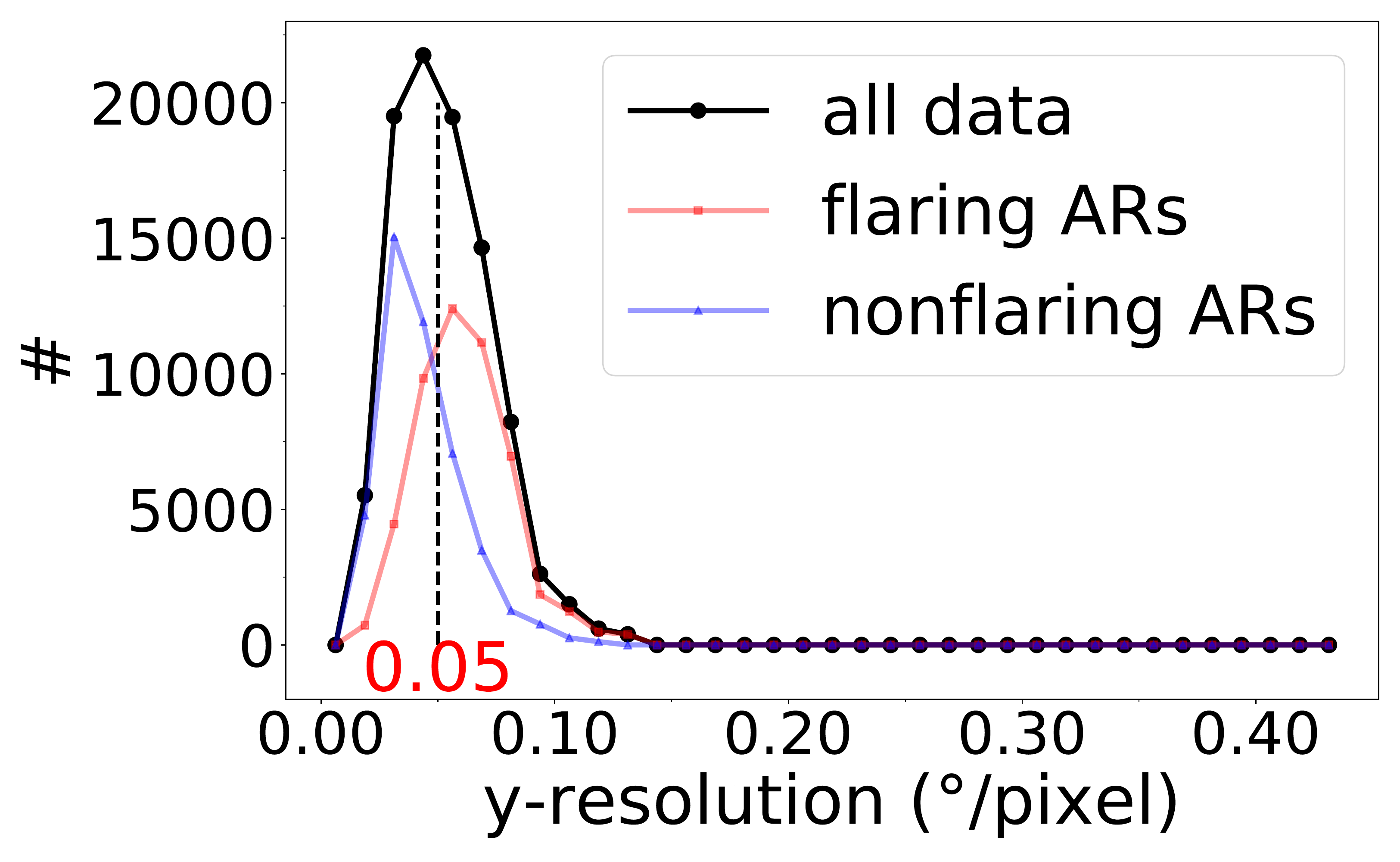}
}
\caption{Distributions of x-resolution ({\it top}) and y-resolution ({\it bottom}) of the HMI/CEA magnetograms resized to $256 \times 256$-pixel images for training the CNN. The x-resolution distribution may be approximated by a Gaussian of width $0.10\degree/{\rm pixel}$ with mean at $0.09\degree/{\rm pixel}$ and the y-resolution may be approximated by a Gaussian of width $0.04\degree/{\rm pixel}$ with mean at $0.05\degree/{\rm pixel}$. Note that as dimensions of the original magnetograms increase, the resolution of the resized images decreases which corresponds to increasing $\degree/{\rm pixel}$.}
\label{fig:SynResDist}
\end{figure}

CNNs typically require input images to be of identical sizes. The line-of-sight magnetograms of flaring and nonflaring ARs used here are, however, varying significantly in size according to AR area . Following \cite{Huang2018}, we resize the AR magnetograms to a fixed size using bi-cubic interpolation. Resizing the magnetograms in this manner yields training images with different spatial resolution depending on the AR area (see Figure \ref{fig:SynResDist}). AR area is known to be a leading factor related to flaring activity \citep{bobraflareprediction,Dhuri2019}. Resizing the images may, therefore, lead to loss of important information about AR area, resulting in sub-optimal CNN classification. Also, convolutional kernels in the CNN are designed to learn spatial features of different length scales from magnetograms, which are important for the classification. The inconsistent spatial resolution of training images hinders CNN kernels from accurately identifying length scales of spatial features correlated with the flaring activity. Keeping in mind these possible drawbacks, we proceed with using resized magnetograms for the CNN classification and later investigate in detail the effect of resizing on CNN performance using synthetic magnetograms (see Section \ref{sec:resSyn}).

We use supervised learning to train CNNs with nonflaring and flaring magnetograms as inputs and labels 0 and 1 respectively as outputs. As per Table \ref{tab:dataset}, we use ARs between May 2010 - Sep 2015 for training and validation of the CNNs. For robust training, we use 10-fold cross-validation as follows. We randomly split the flaring and nonflaring ARs each into three parts and use line-of-sight magnetograms from two parts for training and the remaining part for validation. This process is performed 10 times. Note that all magnetograms of an AR are part of either training or validation set. Also, each flaring and nonflaring AR is sampled $\sim$ 3.3 times on average for the 10-fold cross-validation. After every training, we measure {\it recall} and {\it TSS} for the validation ARs. We tune the hyperparameters --- learning rate $lr$ and minibatch size $N_{\rm BS}$ --- to optimize the mean value of {\it TSS} over the 10 cross-validation runs. We use Python's deep learning library {\it keras} \citep{chollet2015keras} to set up and train the CNNs (see Appendix \ref{app:App1} for details).
\begin{table}[t]
\centering
\begin{tabular}{lcc}
\toprule
& \textbf{CNN-1} & \textbf{CNN-2} \\
\hline
\# Flaring AR images & \multicolumn{2}{c}{10915} \\
\# Nonflaring AR images & \multicolumn{2}{c}{44592}\\
Flaring ARs recall & 0.78  $\pm$ 0.06        & 0.67  $\pm$ 0.06      \\ 
Nonflaring ARs recall & 0.55  $\pm$ 0.04        & 0.78  $\pm$ 0.02      \\ 
TSS          & 0.33 $\pm$ 0.07        & 0.45 $\pm$ 0.07         \\
\hline
\end{tabular}
\caption{10-fold cross-validation performance of CNN-1 (Figure \ref{fig:cnn1}) and CNN-2 (Figure \ref{fig:cnn2}) applied to classify flaring and nonflaring ARs. The AR line-of-sight magnetograms are resized to $128 \times 128$-pixels as input to the CNNs. CNN-2 outperforms the baseline model CNN-1 in terms of the True Skill Statistics ({\it TSS}) score. 1$\sigma$ error bars are determined using 10-fold cross-validation.}    
\label{tab:cnn_comp}
\end{table}

\begin{table}[t]
\centering
\begin{tabular}{lcc}
\toprule
Resized image size & \textbf{128 $\times$ 128} & \textbf{256 $\times$ 256} \\
\hline
Flaring ARs recall & 0.67  $\pm$ 0.06        & 0.71  $\pm$  0.08     \\ 
Nonflaring ARs recall & 0.78  $\pm$ 0.02        & 0.80  $\pm$  0.04       \\ 
TSS          & 0.45 $\pm$ 0.07         & 0.51 $\pm$ 0.06          \\
\hline
\end{tabular}
\caption{10-fold cross-validation performance of CNN-2 (Figure \ref{fig:cnn2}) for the classification of flaring and nonflaring AR line-of-sight magnetograms resized to sizes $128 \times 128$-pixels and $256 \times 256$-pixels. Resizing to $256 \times 256$-pixels yields a higher {\it TSS} score compared to $128 \times 128$-pixels. 1$\sigma$ error bars are obtained using 10-fold cross-validation.}
\label{tab:image_comp}
\end{table}

Since both CNN-1 and CNN-2 require fixed-size inputs, we use magnetograms resized to  $128 \times 128$-pixels for training. From Table \ref{tab:cnn_comp}, we see that CNN-2 yields $\sim~\,10\%$ higher  {\it TSS} at $0.45 \pm 0.07$ than CNN-1. By increasing the size of the resized magnetograms to $256 \times 256$-pixels, {\it TSS} for the classification of flaring and nonflaring ARs increases to $0.51 \pm 0.06$.  {\it TSS} may be increased further by further increasing the size of the resized magnetograms. However, to limit the computational expense, we restrict the analysis to resized AR magnetograms of $256 \times 256$-pixels. Since the CNN-2 10-fold cross-validation performance is significantly higher than that of CNN-1, we use only CNN-2 for the subsequent analysis. CNN-2 yields flaring AR {\it recall} of $0.63 \pm 0.06$, nonflaring AR {\it recall} of $0.89 \pm 0.03$ and classification {\it TSS} of $0.52 \pm 0.04$ on test data comprising ARs between Oct 2015 - August 2018. Note that, because of more severe class imbalance in the test data, the trained CNN performs better in classifying nonflaring ARs and worse in classifying flaring ARs, as expected. The classification TSS, however, is comparable with cross-validation results.
\begin{figure}[b]
\centering
  \includegraphics[width=\columnwidth,trim={4cm 4cm 4cm 3.5cm},clip]{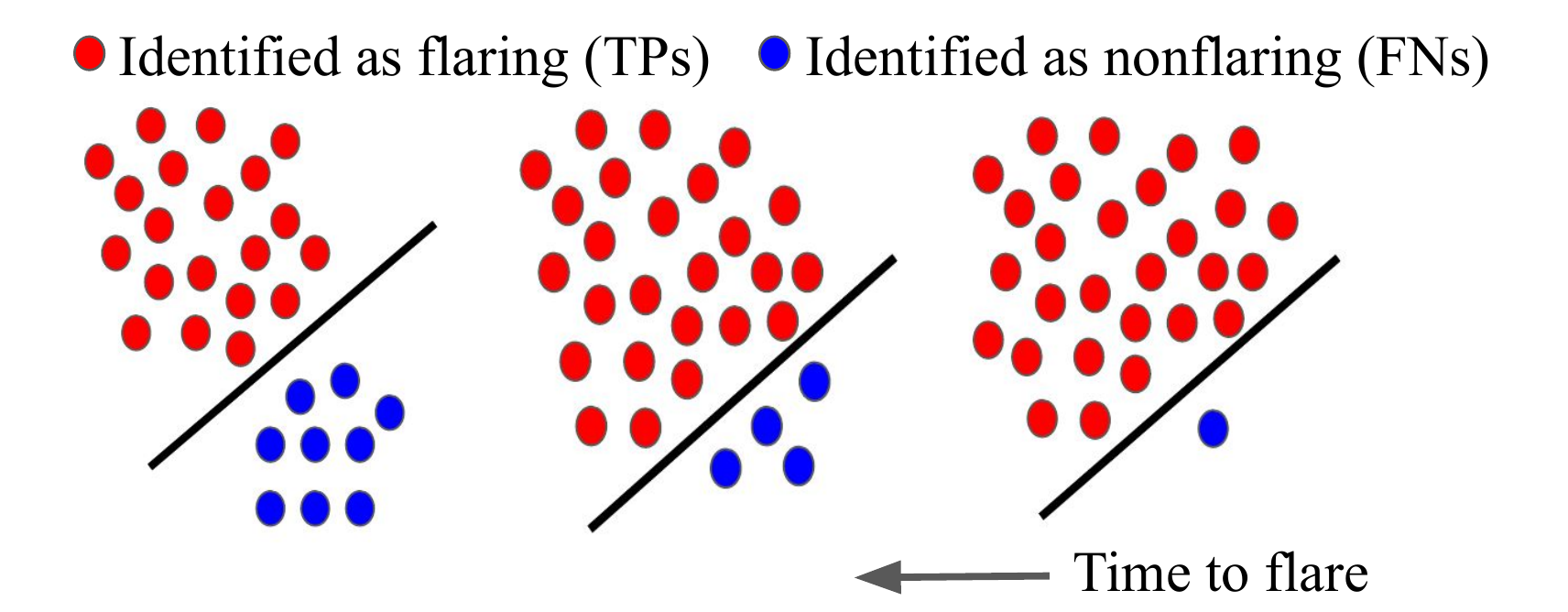}
\caption{Schematic time evolution of the population distribution of line-of-sight magnetogram samples from flaring ARs. As M- or X-class flares approach, the population fraction of True Positive samples ({\it red}) which are accurately identified as flaring is expected to increase and the population fraction of False Negatives ({\it FNs}) which are inaccurately identified as nonflaring is expected to decrease.\label{fig:schemD}}
\end{figure}
\begin{figure*}[t]
\centering
\subfloat{
  \includegraphics[trim = {0cm 0.4cm 0cm 0cm}, clip,width=0.47\textwidth]{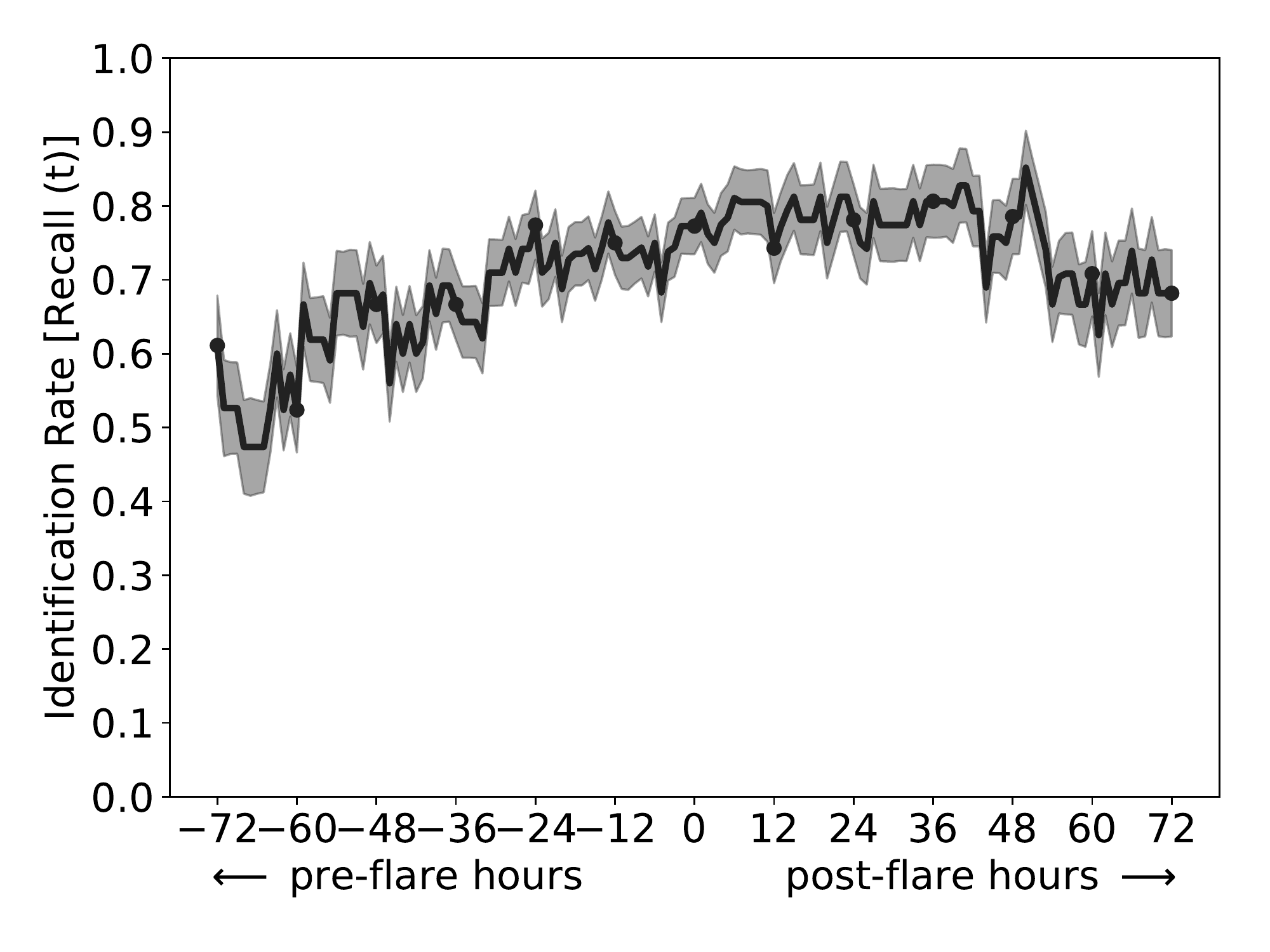}
}
\subfloat{
  \includegraphics[trim = {0cm 0cm 0cm 0.8cm},clip,width=0.48\textwidth]{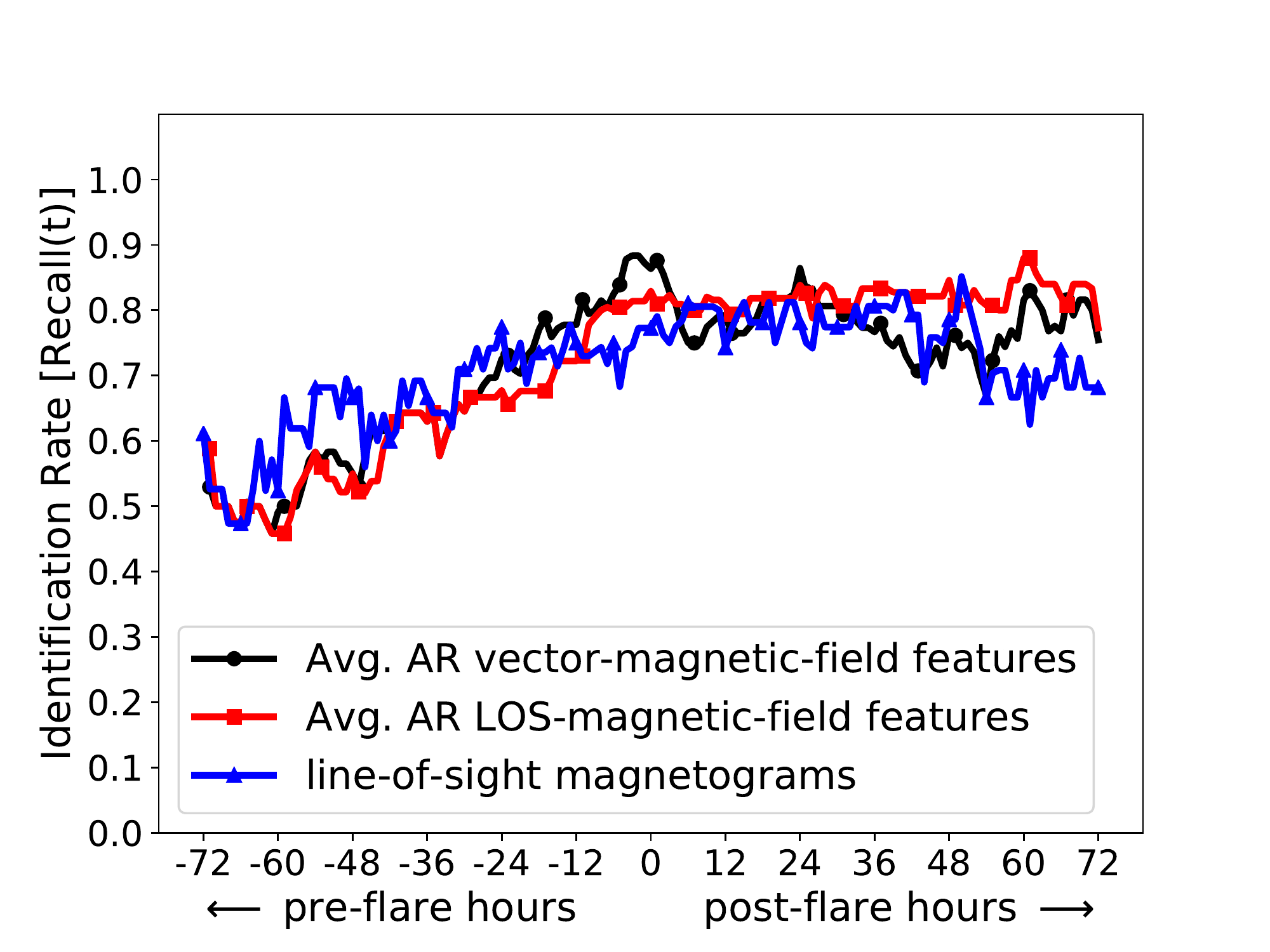}
}

\caption{{\it Left panel:} Time evolution of instantaneous identification rate or instantaneous $recall(t_r) = TP(t_r)/(TP(t_r) + FN(t_r))$ as predicted by the CNN for the validation data. Here, $t_r$ represents time relative to the flare event. The instantaneous {\it recall} is high $> 50\%$ for days before and after M- or X-class flares. The instantaneous {\it recall} rises from a value of $\sim 50\%$, 72 hours before the flares and peaks at $\sim 80\%$. The shaded area represents 1$\sigma$ error. {\it Right panel:} Comparison of the instantaneous {\it recall} of the CNN ({\it blue}) with a Support Vector Machine trained \citep{Dhuri2019} using AR averaged vector-magnetic-field SHARP features \citep{Bobra2014} ({\it black})  as well as SHARP features that can be reliably interpreted from line-of-sight magnetograms viz. AR area and the total unsigned flux ({\it red}). The SVM trained on full vector-magnetic-field SHARP features outperforms the CNN. The performance of the CNN is comparable to the SVM trained with SHARPs that may be inferred from line-of-sight magnetograms. \label{fig:instR}}
\end{figure*}

\subsection{Machine correlation between line-of-sight magnetograms and flaring activity}
\begin{figure*}[t]
\centering
\subfloat{
  \includegraphics[clip,width=0.45\textwidth]{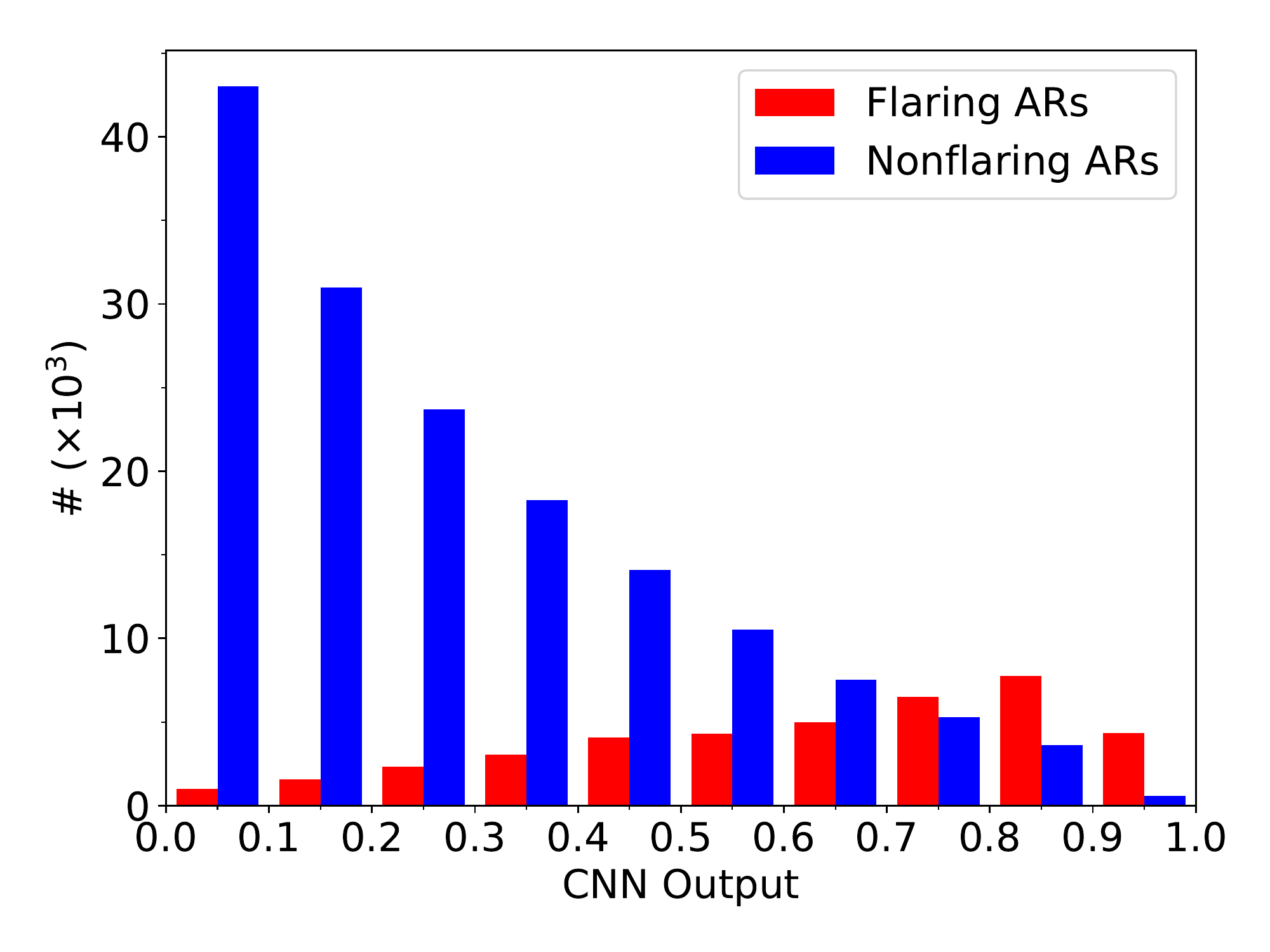}
}
\subfloat{
  \includegraphics[clip,width=0.45\textwidth]{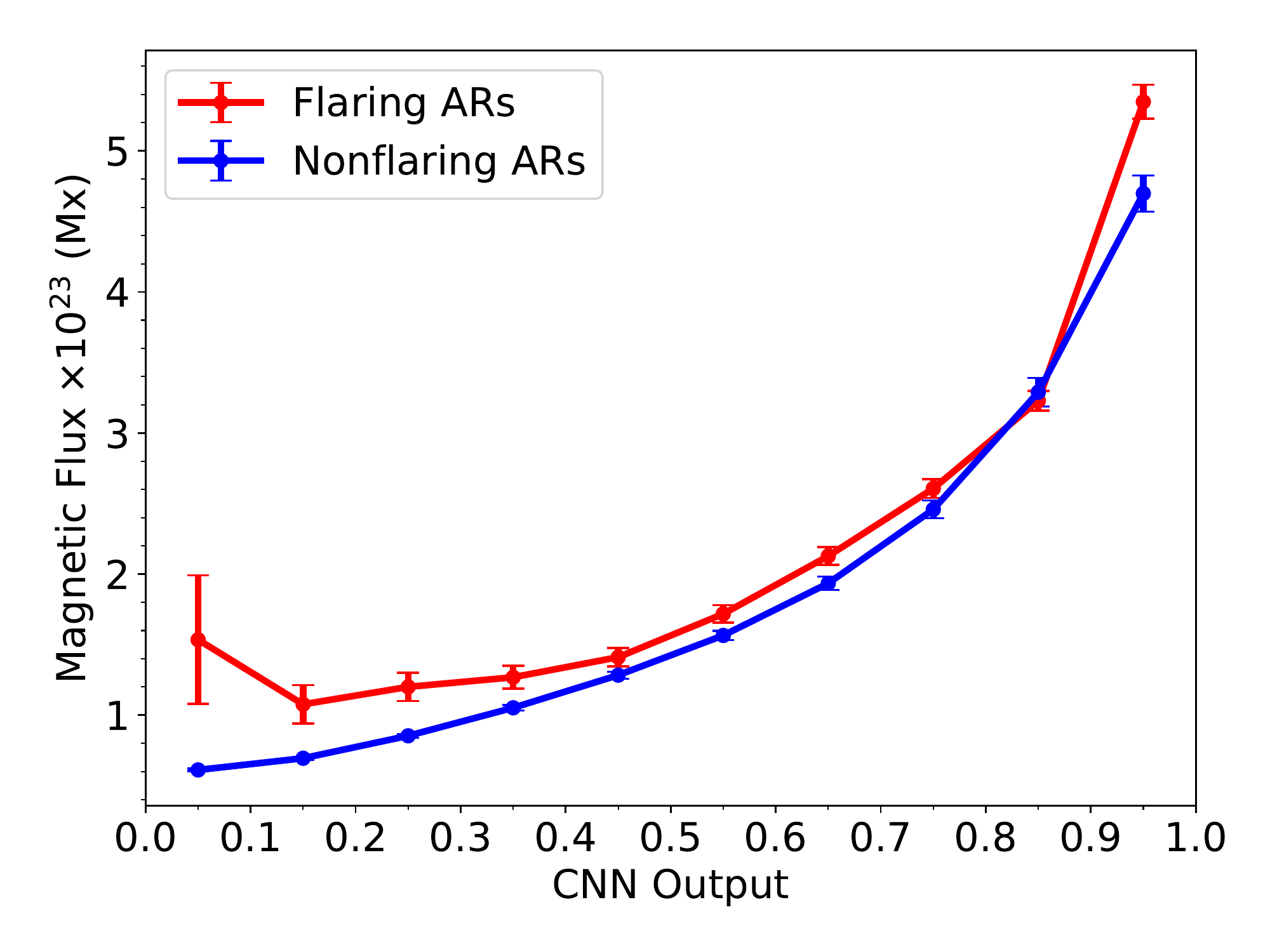}
}
\caption{Statistical analysis of the total unsigned line-of-sight magnetic flux of flaring and nonflaring ARs. The flaring and nonflaring line-of-sight magnetograms are categorized in bins depending on the corresponding CNN output. These bins are centered at $y=\{0.05,0.15,0.25,0.35,0.45,0.55,0.65,0.75,0.85,0.95\}$ and bounded by $y \pm \Delta/2$ where $\Delta=0.1$. {\it Left panel:} the histogram of flaring and nonflaring samples binned by the CNN output. For a significant majority of nonflaring AR samples, the CNN output $y < 0.5$ and for a significant majority of flaring AR samples $y > 0.5$. {\it Right panel:} The average unsigned line-of-sight magnetic flux calculated for flaring and nonflaring magnetograms from each bin of the CNN output. The average unsigned line-of-sight magnetic flux for both flaring and nonflaring samples increases as the CNN output increases. $5\sigma$ error bars are shown. \label{fig:SA}}
\end{figure*}
The trained CNN-2, henceforth referred to as the CNN, classifies between the line-of-sight magnetograms of flaring and nonflaring ARs with {\it TSS} of $\sim 50\%$. Note that the classification {\it TSS} is indicative of the success in identifying a magnetogram from flaring ARs irrespective of the observation time relative to flare. We expect that the CNN identification is better for magnetograms which are observed a few hours before flares compared to those that are observed well away from the flare event (Figure \ref{fig:schemD}). Thus, we expect that the population fraction of accurately identified line-of-sight magnetograms increases as flare time approaches. A measure of the instantaneous population fraction of accurately identified magnetograms from flaring ARs is the {\it recall} or identification rate calculated as $recall(t_r) = TP(t_r) / \left(TP(t_r) + FN(t_r)\right)$, where $t_r$ is time relative to M- or X-class flares. The instantaneous {\it recall} or identification rate can be interpreted as a correlation of line-of-sight magnetograms with flaring activity calculated using the CNN. Time evolution of the instantaneous {\it recall} is thus indicative of dynamics of the line-of-sight magnetic fields before and after flares. To calculate the instantaneous {\it recall}, we compile time series of magnetograms from flaring ARs during a window $t_r=t-T_F\in[-72,72] \textrm{ hours}$ centered around a flare event $T_F$. If two consecutive flares on an AR are separated by $< 144$ hours, we split the observations between the flare events in two halves and consider the first half as the post-flare category of the first flare and the second half as the pre-flare category of the second flare. We align all such time series from flaring ARs at $t-T_F=t_r=0$, the time of flare events. Using all the aligned time series of magnetograms, we obtain the instantaneous {\it recall} for flaring AR magnetograms within $\pm 72$ hours of flares. 

{\it Left panel} of the Figure \ref{fig:instR} shows the temporal evolution of the instantaneous {\it recall} of magnetograms from flaring ARs in validation data from 72 hours before flares to 72 hours after. We find that the instantaneous {\it recall} is $> 0.5$ for days before and after flares. This suggests that flaring ARs remain in a flare-productive state for days before and after flares. The instantaneous {\it recall} peaks at $\sim 0.8$, which is consistent with reported results for flare forecasting \citep{Huang2018}. \cite{Dhuri2019} obtained the instantaneous {\it recall} using a support vector machine (SVM) trained on the AR averaged vector-magnetic-field features viz. SHARP features \citep{bobraflareprediction}. The {\it right panel} of Figure \ref{fig:instR} compares the CNN trained on line-of-sight magnetograms and the SVM trained on AR-averaged vector-magnetic-field features. We find that the peak SVM instantaneous {\it recall} is $\sim 10\%$ higher than the CNN trained on line-of-sight magnetograms. Also, the SVM trained using AR-averaged features that may be inferred from the line-of-sight magnetograms --- namely AR area and total unsigned flux --- shows performance approximately the same as the CNN. This suggests that the CNN output largely depends on the AR area and total unsigned magnetic flux.
\begin{figure}[b]
\centering
  \includegraphics[trim = {0.4cm 1.4cm 0.7cm 1.4cm},clip,width=0.3\textwidth]{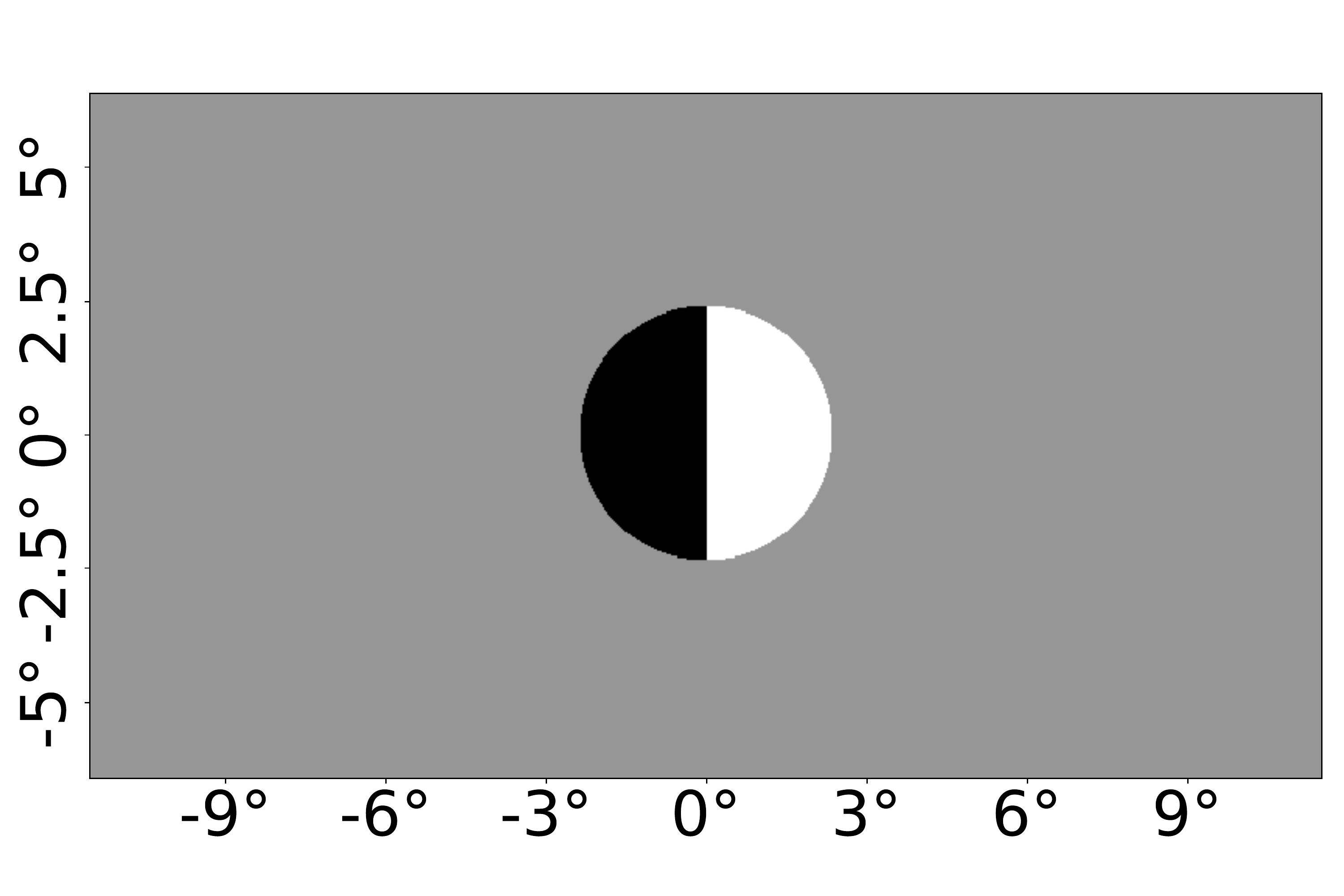}
\caption{A $23\degree \times 12.8\degree$ magnetogram ($728 \times 427$-pixels) of a synthetic bipole with uniform field used for probing the CNN. The synthetic bipole has $-+$ configuration (indicated by {\it black} and {\it white} respectively) and the magnetic field strength is 2000\,G.}
\label{fig:SynMag}
\end{figure}

\subsection{Statistical analysis of the CNN output}
From the instantaneous {\it recall} in Figure \ref{fig:instR}, the CNN output seems to primarily depend upon the total unsigned magnetic flux of flaring and nonflaring ARs. Therefore, we perform statistical analysis of the total unsigned line-of-sight magnetic flux of flaring and nonflaring magnetograms categorized according to CNN output. Flaring and nonflaring magnetograms are binned into ten buckets based on the associated CNN output as shown in Figure \ref{fig:SA}. The {\it left panel} of Figure \ref{fig:SA} shows the number of magnetograms categorized according to the bins of the CNN output. Note that magnetograms for which the CNN output $y\geq0.5$ are identified as flaring and $y<0.5$ are identified as nonflaring. We see that, for a significant number of nonflaring ARs, the CNN output $y \sim 0.0$ and for a significant number of flaring ARs, $y \sim 1.0$. The {\it right panel} of Figure \ref{fig:SA} displays the average value of total unsigned line-of-sight magnetic flux for flaring and nonflaring magnetograms from each bin of the CNN output. The average value of the total unsigned line-of-sight magnetic flux systematically increases for both flaring as well as nonflaring AR magnetograms as a function of the CNN output. Thus, the CNN output is highly correlated with the total unsigned line-of-sight magnetic flux.
\begin{figure*}[t]
\centering
\subfloat{
  \includegraphics[clip,width=0.35\textwidth]{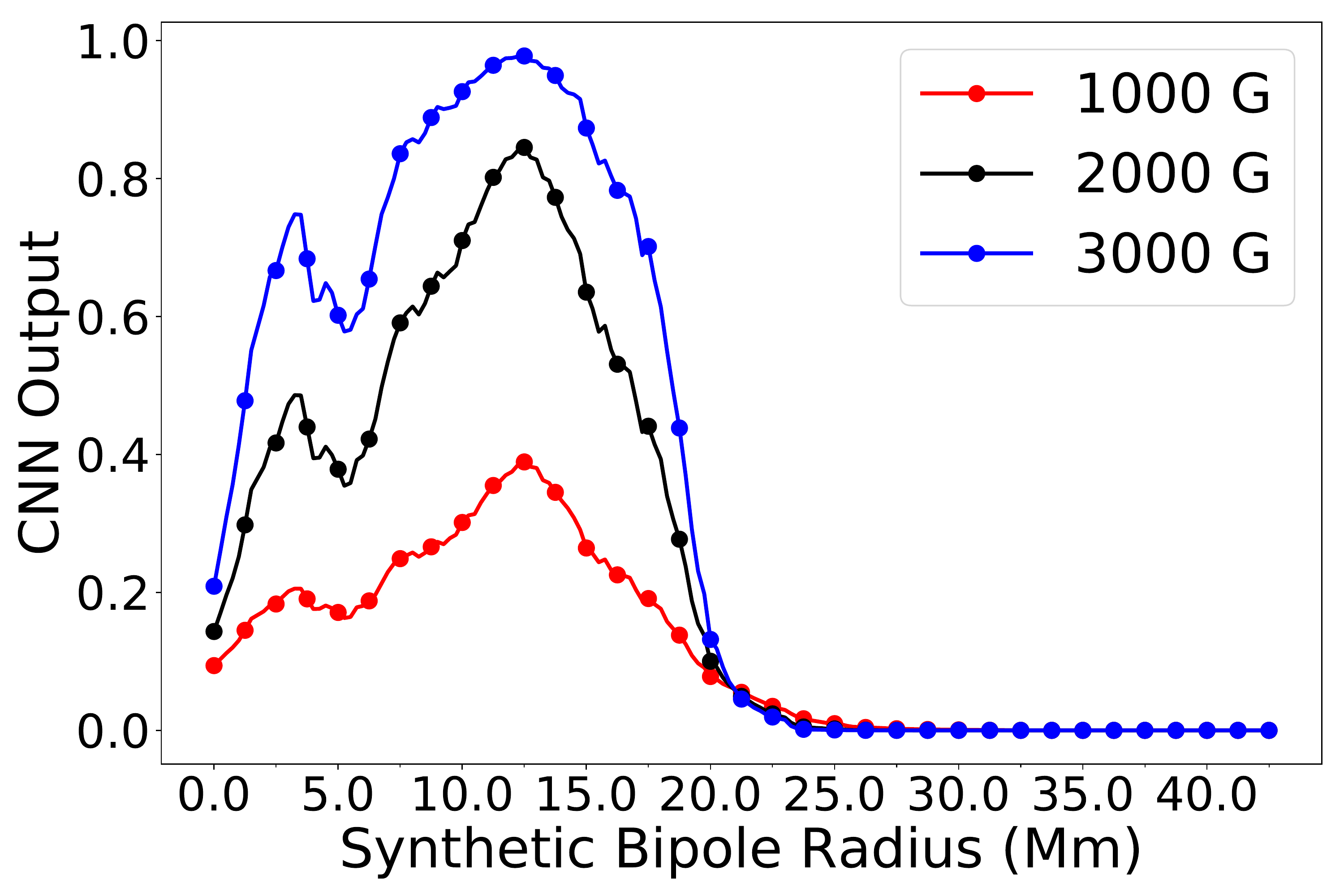}
}
\subfloat{
  \includegraphics[clip,width=0.35\textwidth]{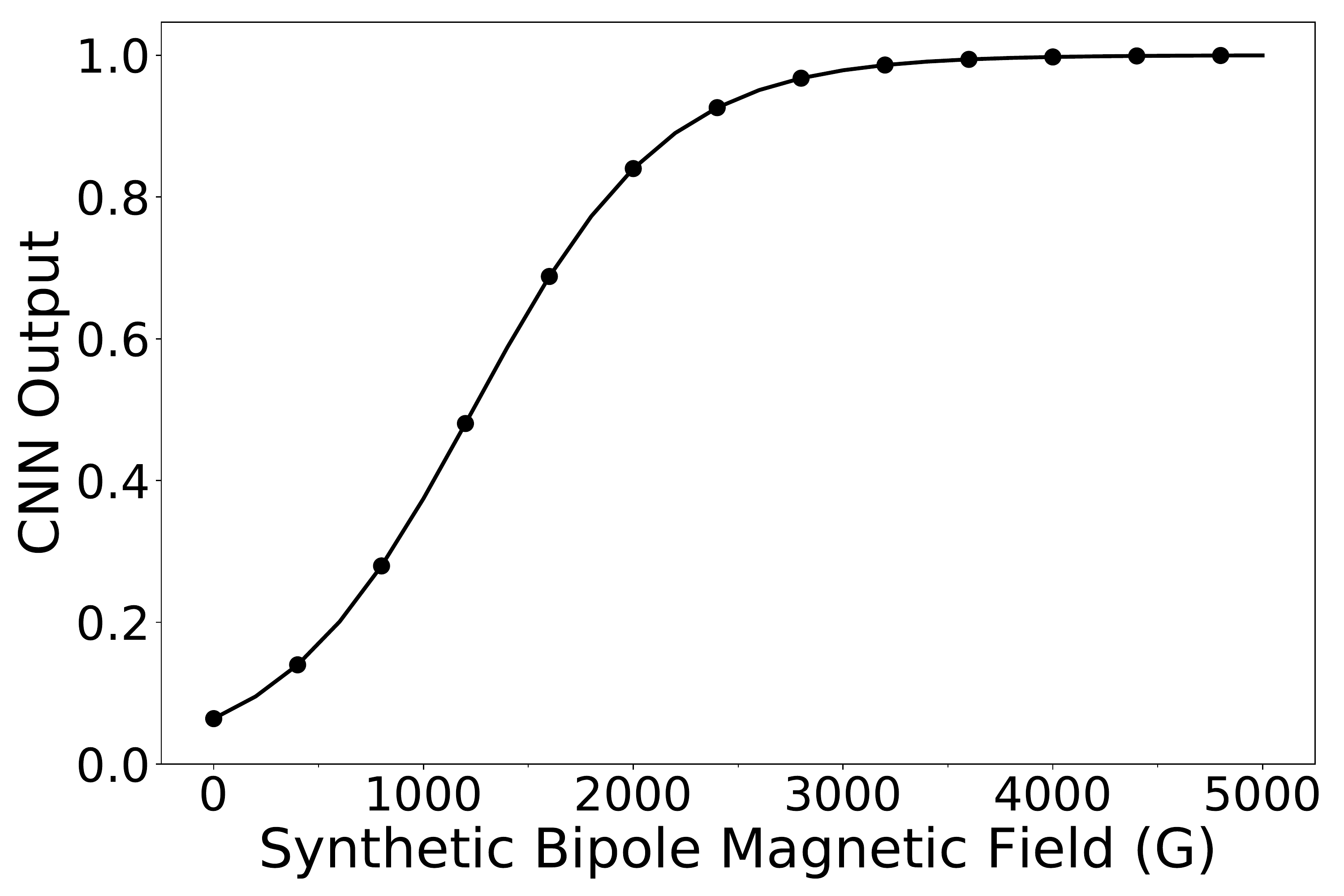}}\\
\subfloat{
  \includegraphics[clip,width=0.35\textwidth]{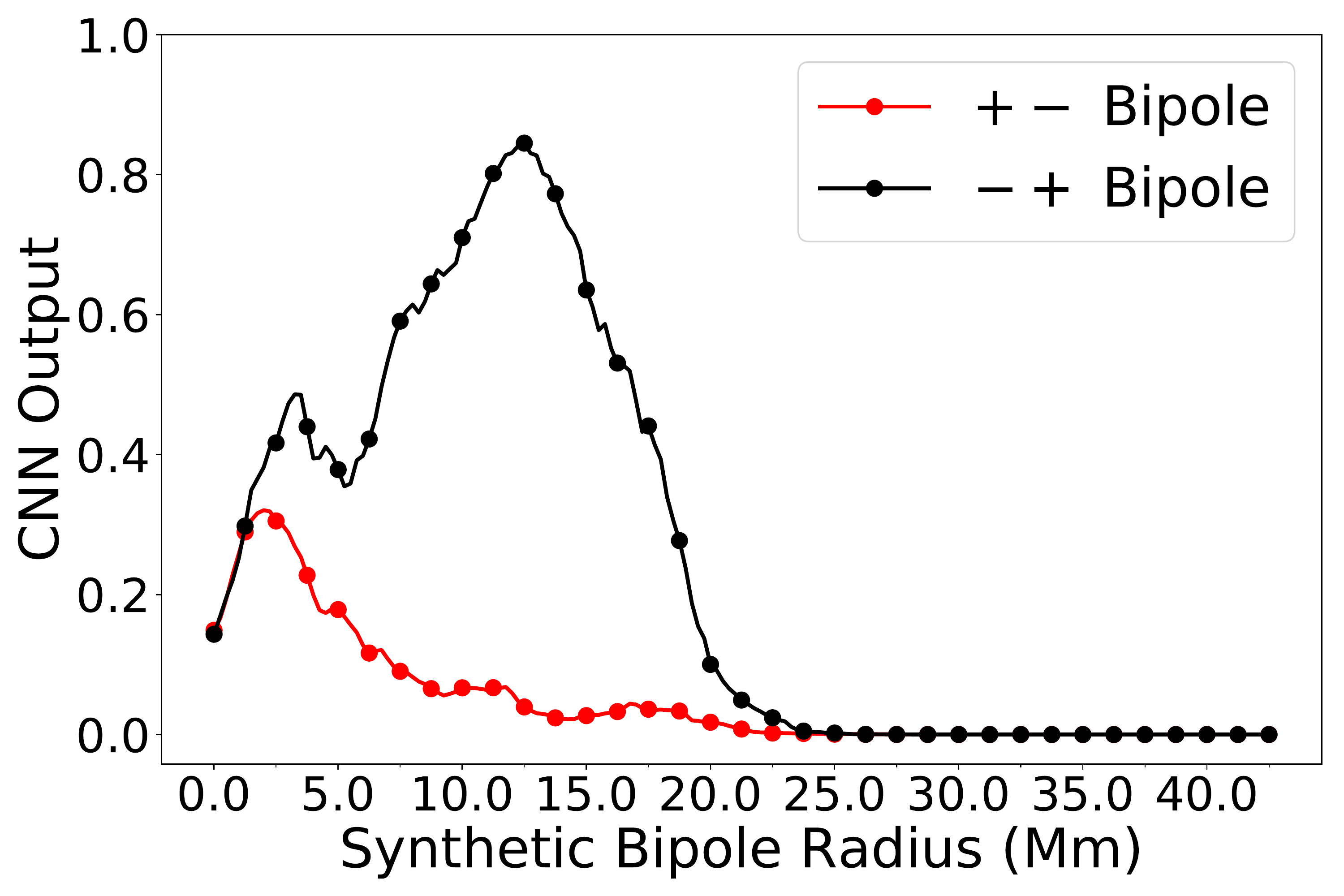}}
\subfloat{
  \includegraphics[clip,width=0.35\textwidth]{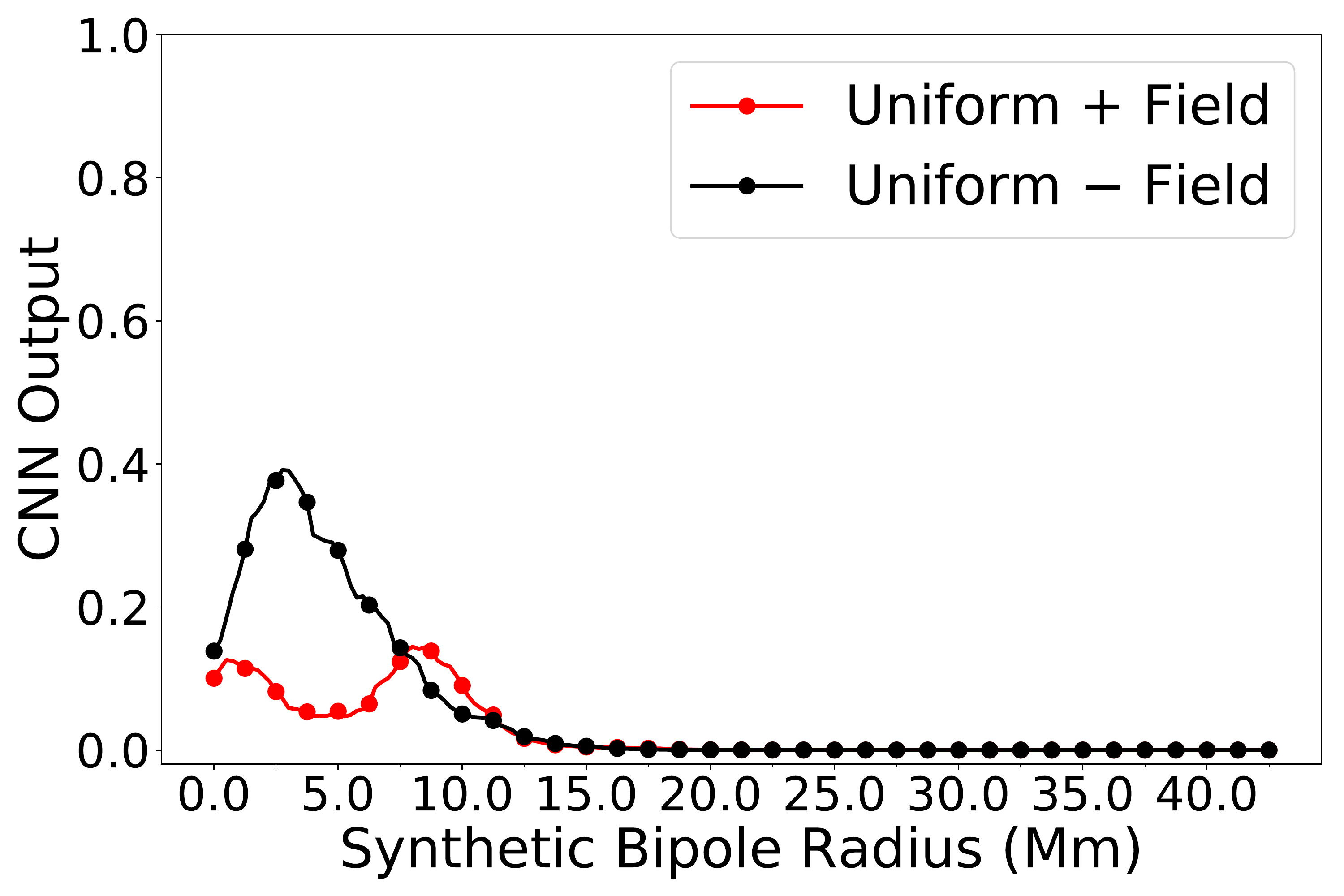}}
\caption{Probing the trained CNN with synthetic bipoles on $768 \times 427$-pixel magnetograms which yield resized images with ${\rm x-res.}=0.09\degree/{\rm pixel}$ and ${\rm y-res.}=0.05\degree/{\rm pixel}$. {\it Top left:} Variation of the CNN output with the radius of the $-+$ synthetic bipole for uniform field strengths of 1000\,G, 2000\,G, and 3000\,G. CNN output curves show a low peak at $\sim {\rm 3.5\,Mm}$ and a high peak at $\sim {\rm 12.5\,Mm}$. The output increases with field strength and falls rapidly as the radius of the synthetic bipole increases beyond $\sim 20\,{\rm Mm}$. {\it Top right:} Variation of CNN output as a function of field strength for a $-+$ configuration synthetic bipole of radius $\sim {\rm 12.5\,Mm}$. The output increases with increasing field strength and asymptotically approaches 1 for fields $>3000\,{\rm G}$. {\it Bottom left:} Dependence of CNN output on the configuration of the synthetic bipole --- $+-$ configuration and $-+$ configuration --- with a uniform field of 2000\,G. The CNN output is higher for a $-+$ configuration bipole than the $+-$ configuration bipole of the same size. {\it Bottom right:} Variation of CNN output with size for a circular magnetic region of the uniform field of 2000\,G. The output for a magnetic region with uniform negative ($-$) field is higher than the magnetic region of the same size with uniform positive ($+$) field.}
\label{fig:SynP}
\end{figure*}
\begin{figure*}[t]
\centering
\subfloat{
  \includegraphics[clip,width=0.35\textwidth]{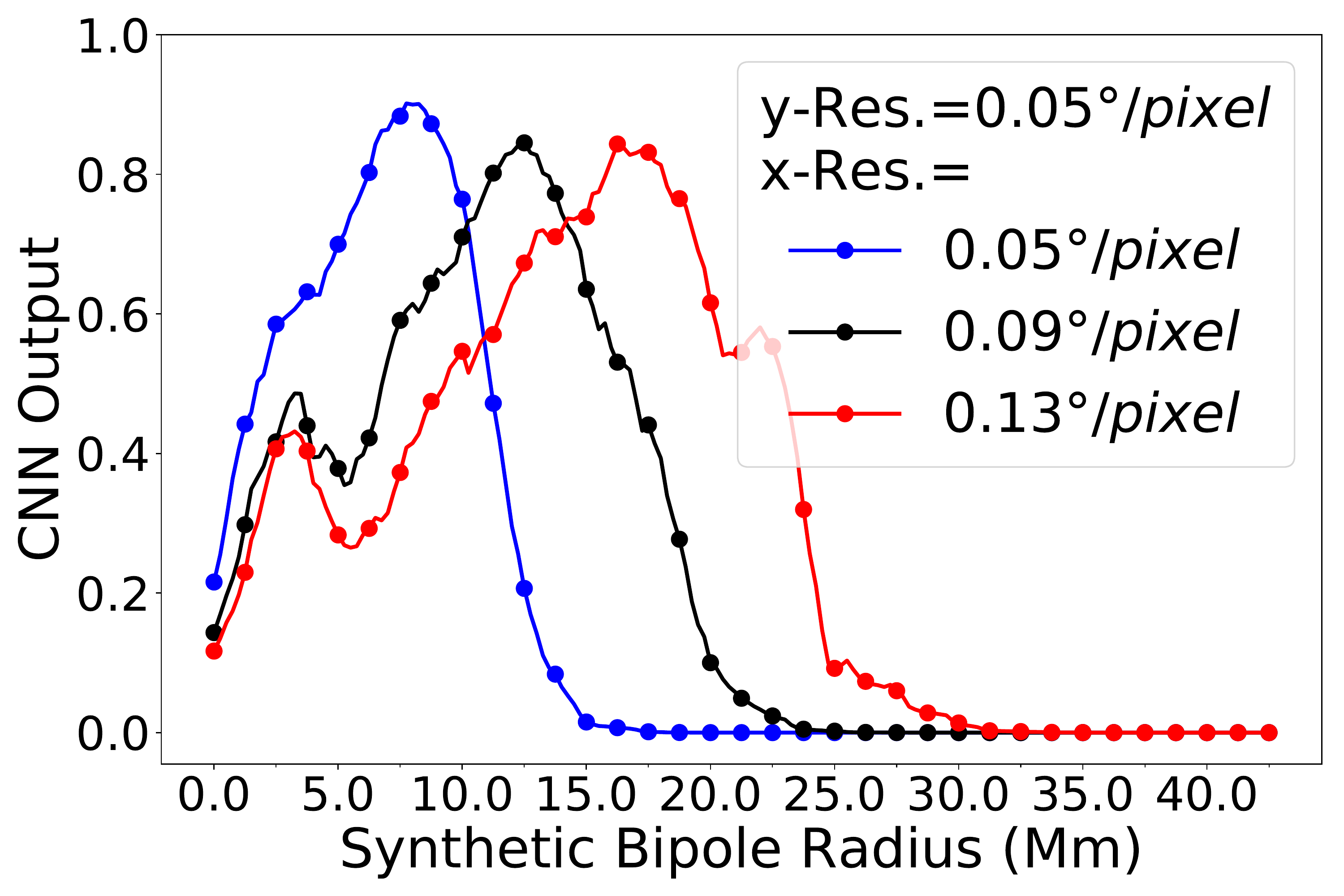}}
\subfloat{
  \includegraphics[clip,width=0.35\textwidth]{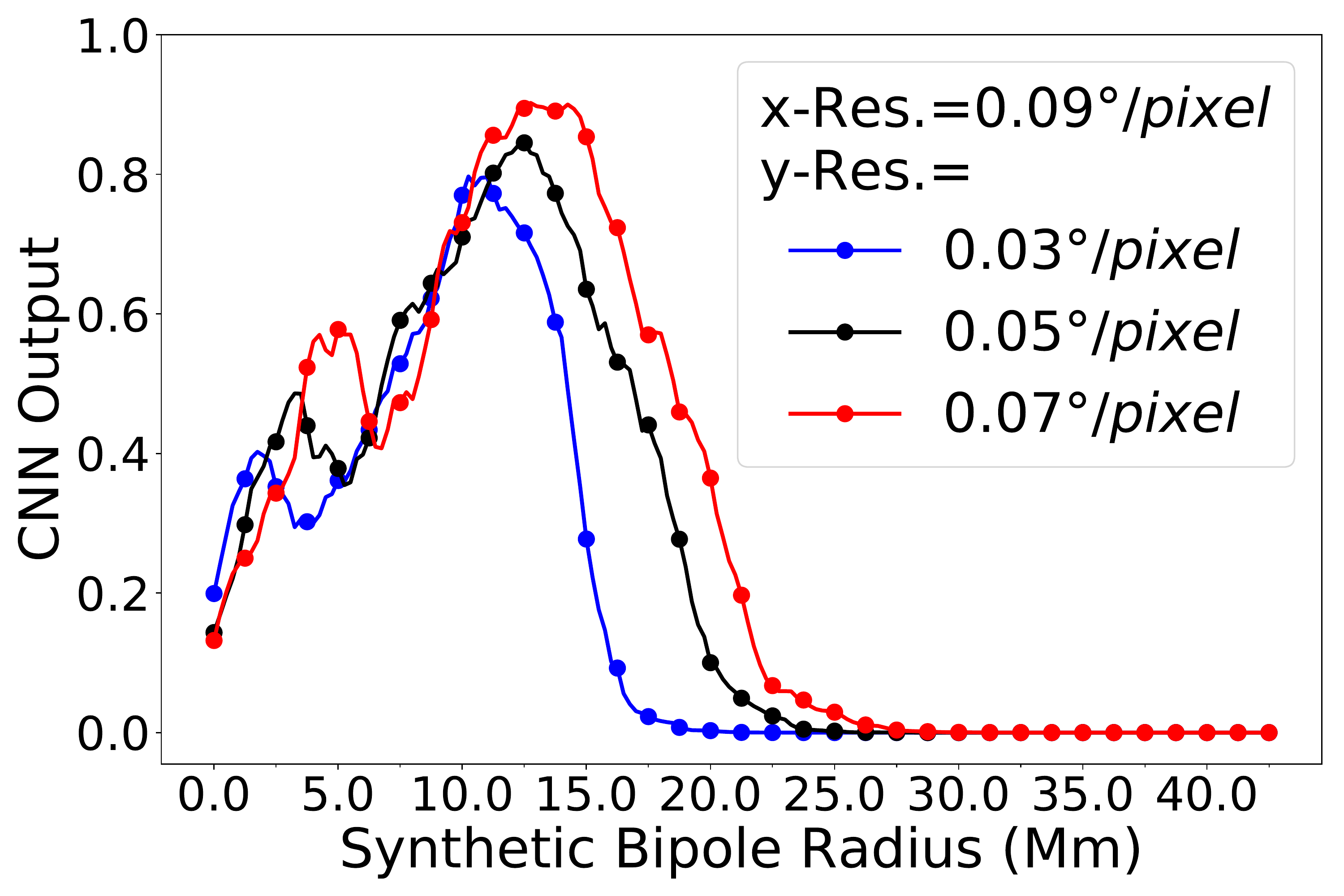}}
\subfloat{
  \includegraphics[clip,width=0.24\textwidth]{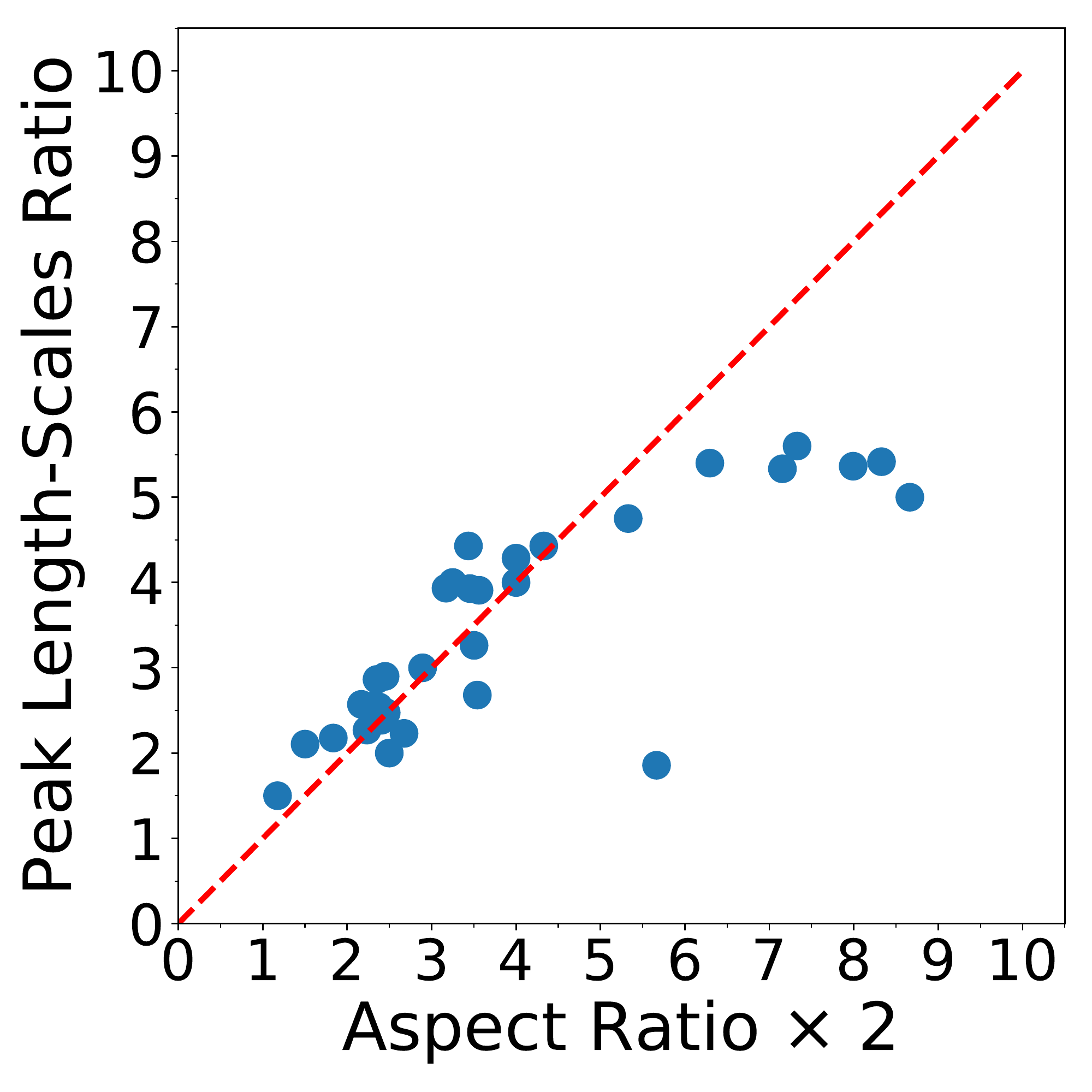}}
\caption{Dependence of the CNN output on the magnetogram dimensions. {\it Left:} The CNN output variation with the radius of the synthetic bipole with $-+$ configuration for magnetograms which yield resized images of ${\rm x-res.} = 0.05\degree/{\rm pixel},\ 0.09\degree/{\rm pixel},\ {\rm and} \ 0.13\degree/{\rm pixel}$ with fixed ${\rm y-res.} = 0.05\degree/{\rm pixel}$. The length-scale corresponding to high peak increases as the x-resolution decreases. {\it Center:} The CNN output variation with the radius of the synthetic bipole with $-+$ configuration for magnetograms which yield resized images of ${\rm y-res.} = 0.03\degree/{\rm pixel},\ 0.05\degree/{\rm pixel},\ {\rm and}\ 0.07\degree/{\rm pixel}$ with fixed ${\rm x-res.} = 0.09\degree/{\rm pixel}$. The length-scale corresponding to low peak increases as the y-resolution decreases. {\it Right:} Correlation between the ratio of length-scales corresponding to high and low peaks with the aspect ratio (x-dimension/y-dimension) of the magnetograms. The ratio of the length-scales at high and low peaks is approximately two times the aspect ratio.}
\label{fig:SynRes}
\end{figure*}
\begin{figure}[t]
\centering
\subfloat{
  \includegraphics[trim = {0.2cm 0.2cm 0.2cm 0.2cm},clip,width=0.35\textwidth]{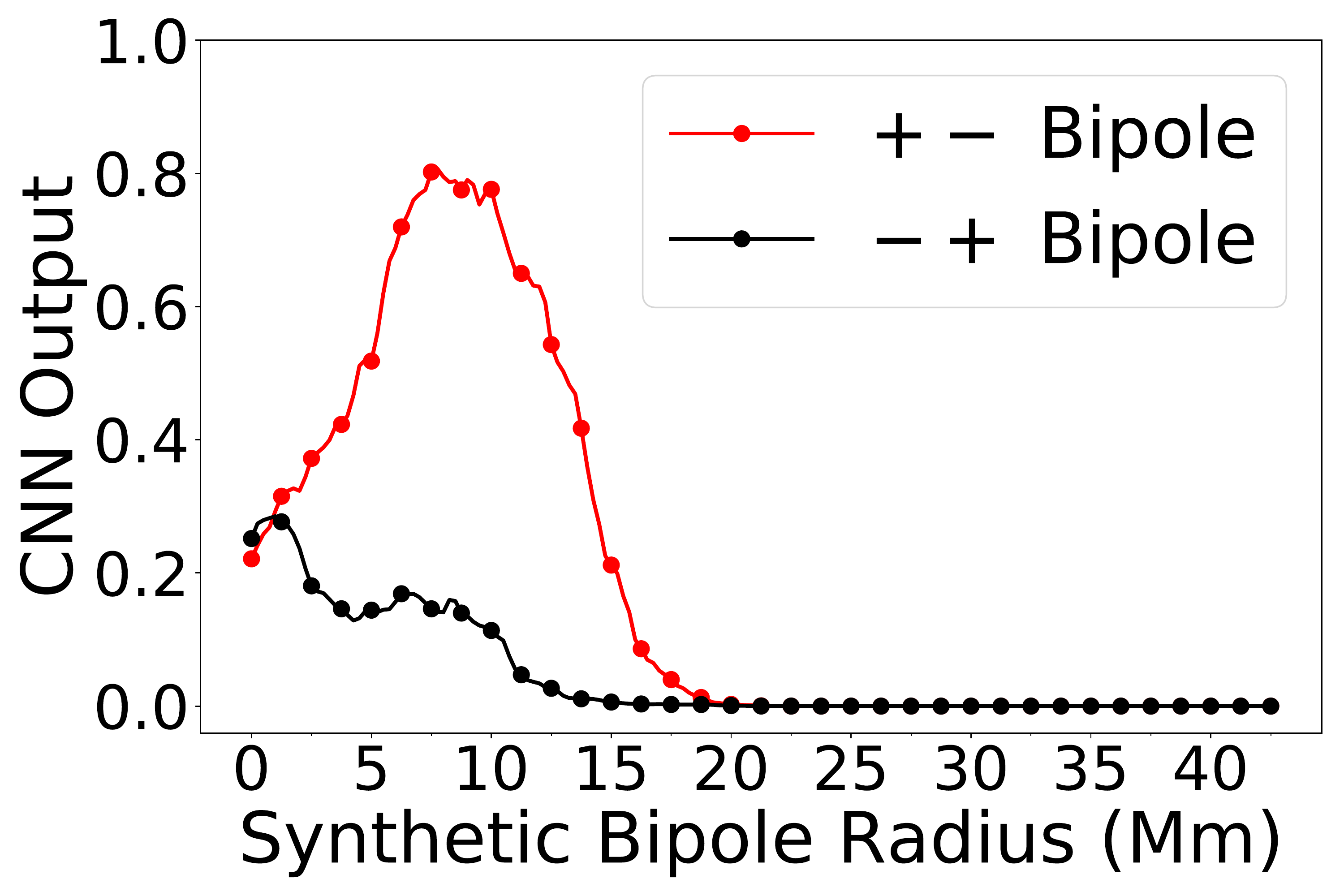}
}\\
\subfloat{
  \includegraphics[trim = {0.2cm 0.2cm 0.2cm 0.2cm},clip,width=0.35\textwidth]{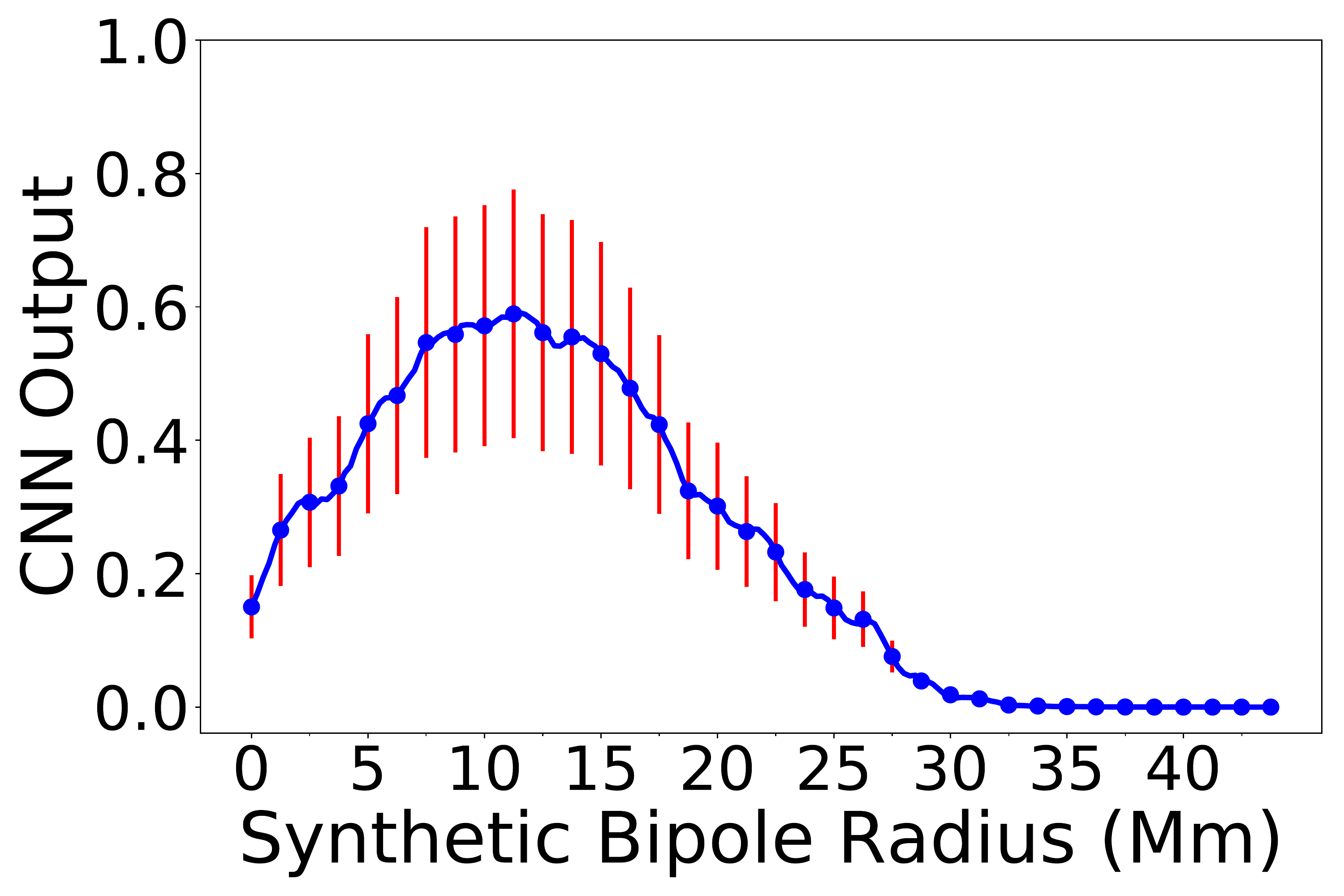}
}
\caption{{\it Top:} Dependence of the CNN output on the configuration of the synthetic bipole --- $+-$ configuration and $-+$ configuration --- with a uniform field of 2000\,G after reversing the training magnetograms' polarity. {\it Bottom:} The variation of the mean CNN output of 10-cross-validation models with the size of the synthetic bipole of $-+$ configuration and a uniform field of 2000\,G. $1\sigma$ standard error is shown.}
\label{fig:SynConfg}
\end{figure}
\begin{figure*}[t]
\centering
\subfloat{
  \includegraphics[trim = {1cm 6cm 1cm 4cm},clip,width=0.5\textwidth]{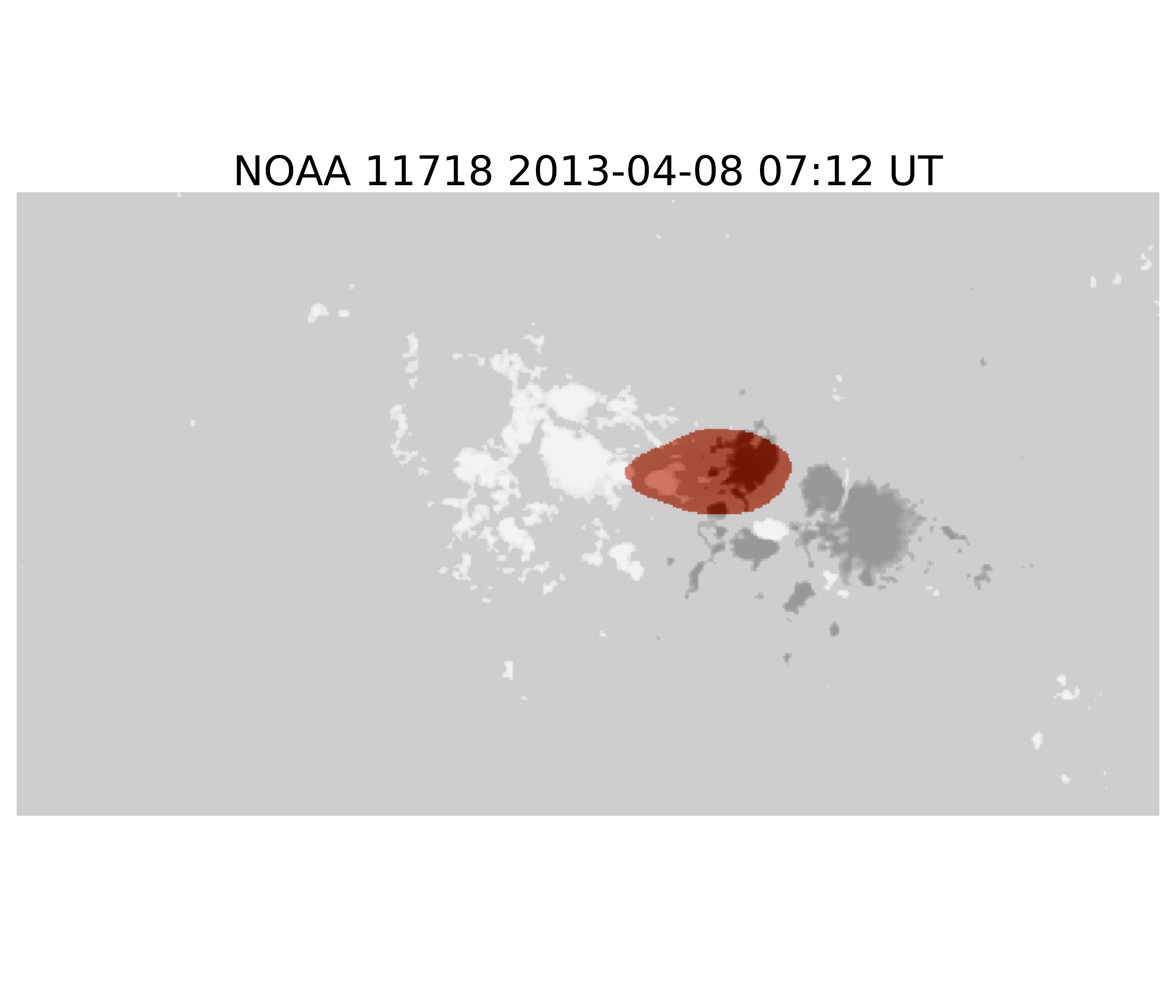}
}
\subfloat{
  \includegraphics[trim = {1cm 6cm 1cm 4cm},clip,width=0.5\textwidth]{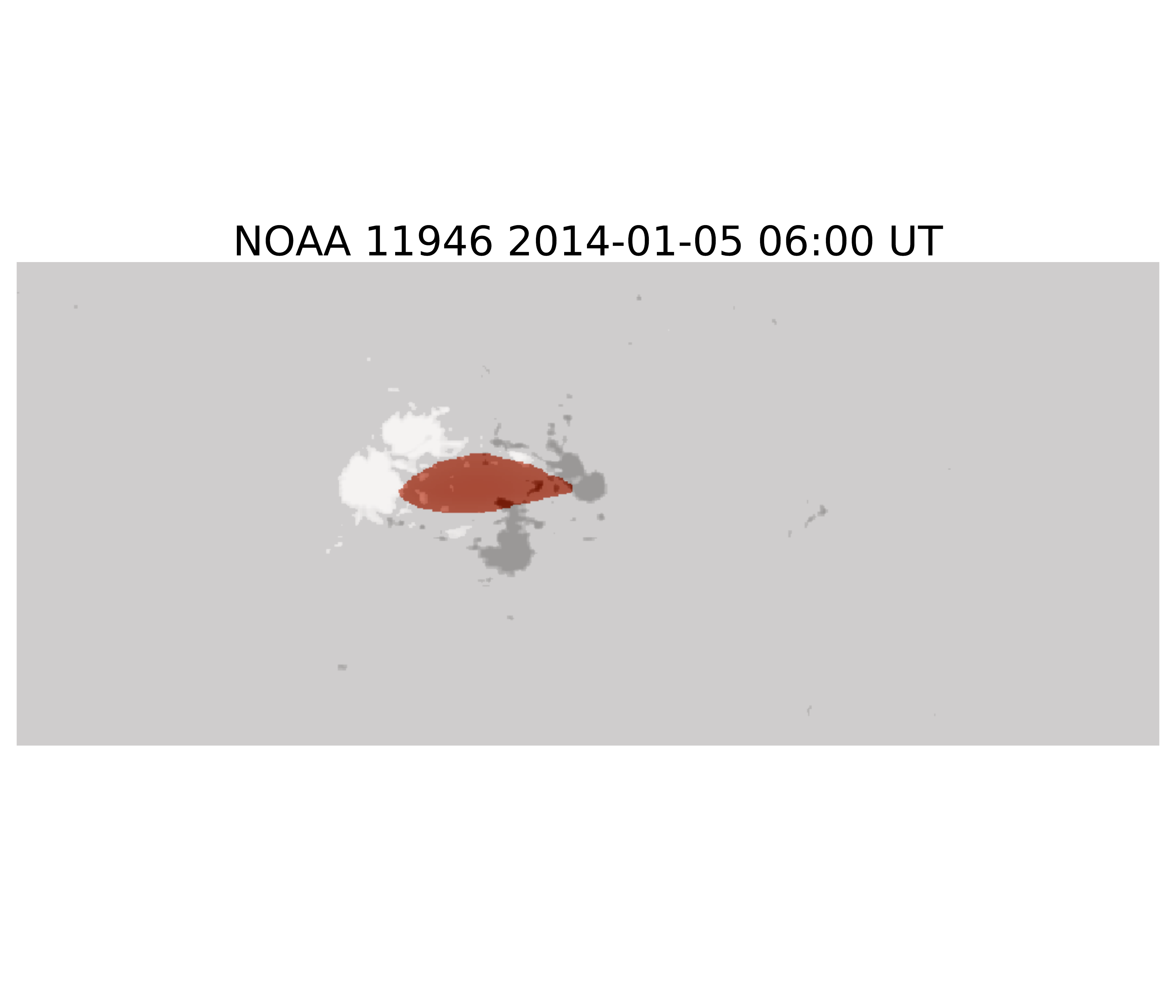}
}\\
\subfloat{
  \includegraphics[trim = {1cm 6cm 1cm 4cm},clip,width=0.5\textwidth]{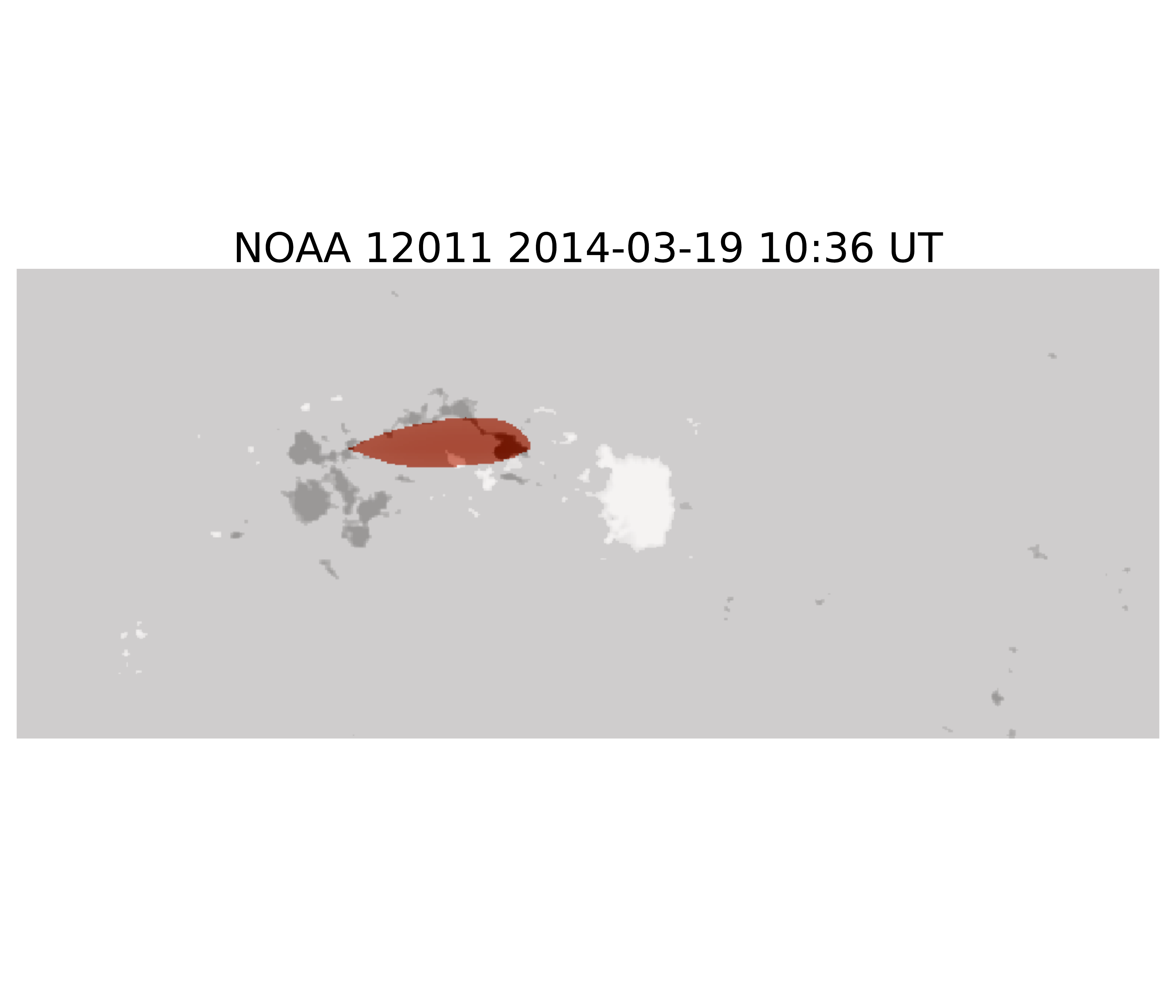}}
\subfloat{
  \includegraphics[trim = {1cm 6cm 1cm 4cm},clip,width=0.5\textwidth]{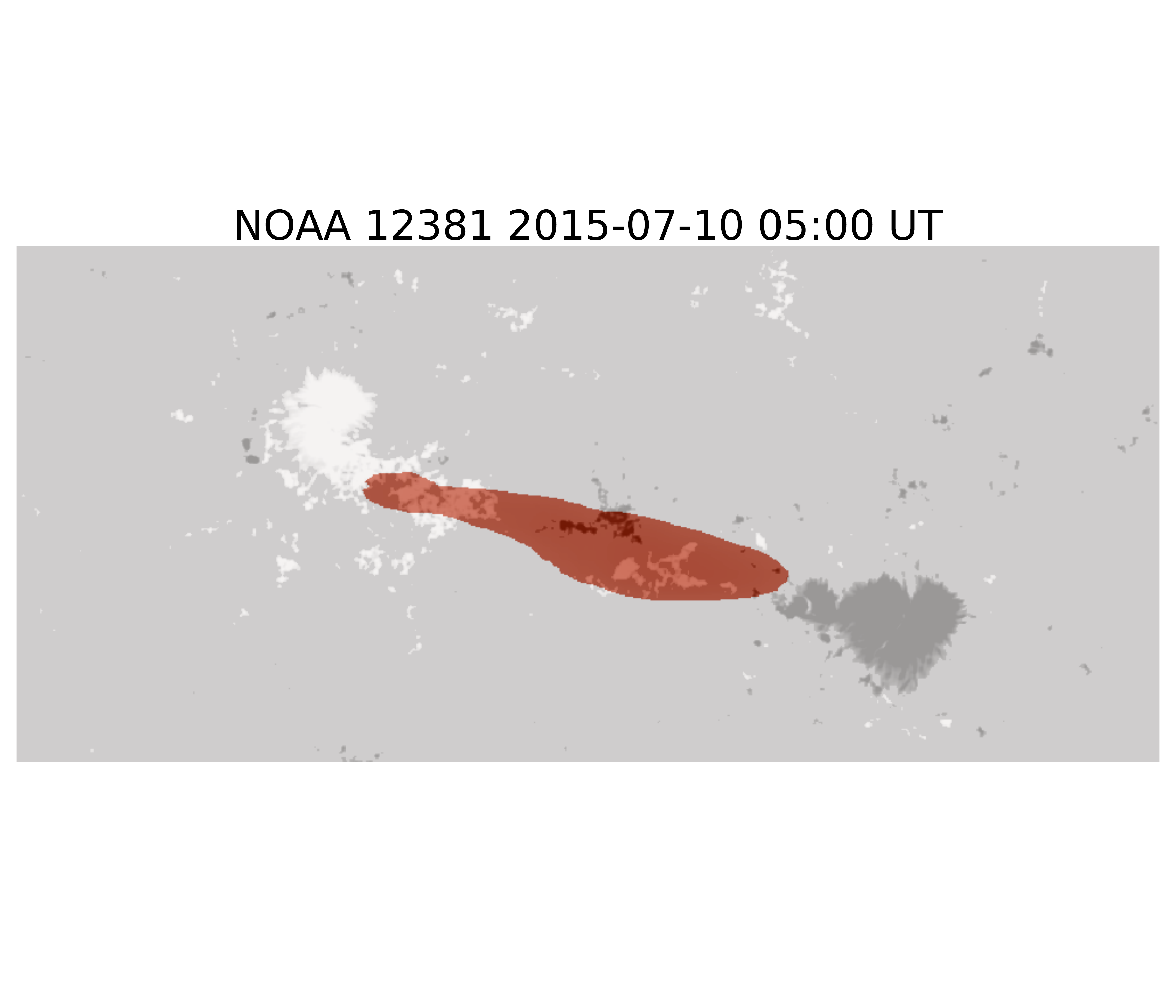}}\\
\subfloat{
  \includegraphics[trim = {1cm 1cm 1cm 0cm},clip,width=0.5\textwidth]{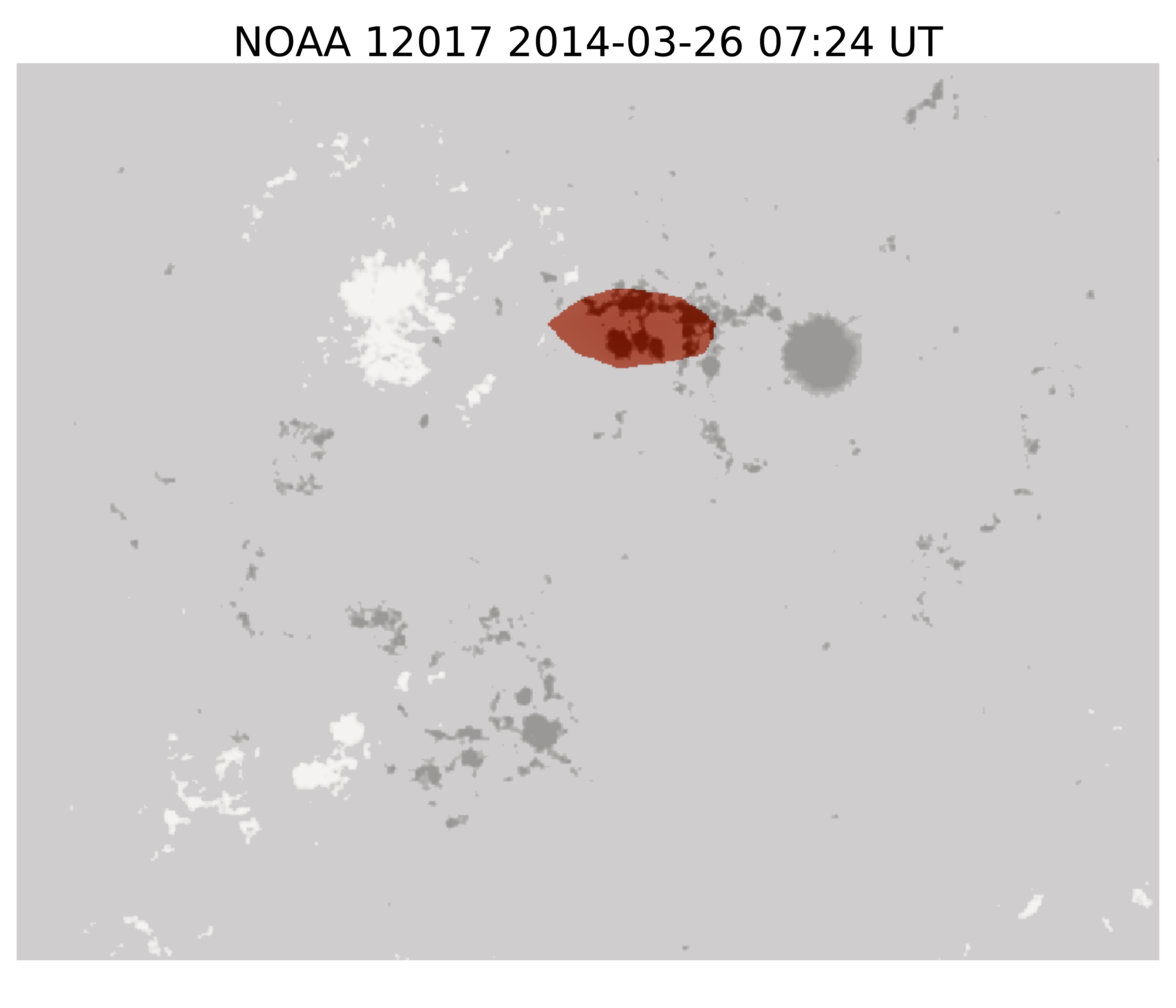}}
\subfloat{
  \includegraphics[trim = {1cm 1cm 1cm 0cm},clip,width=0.5\textwidth]{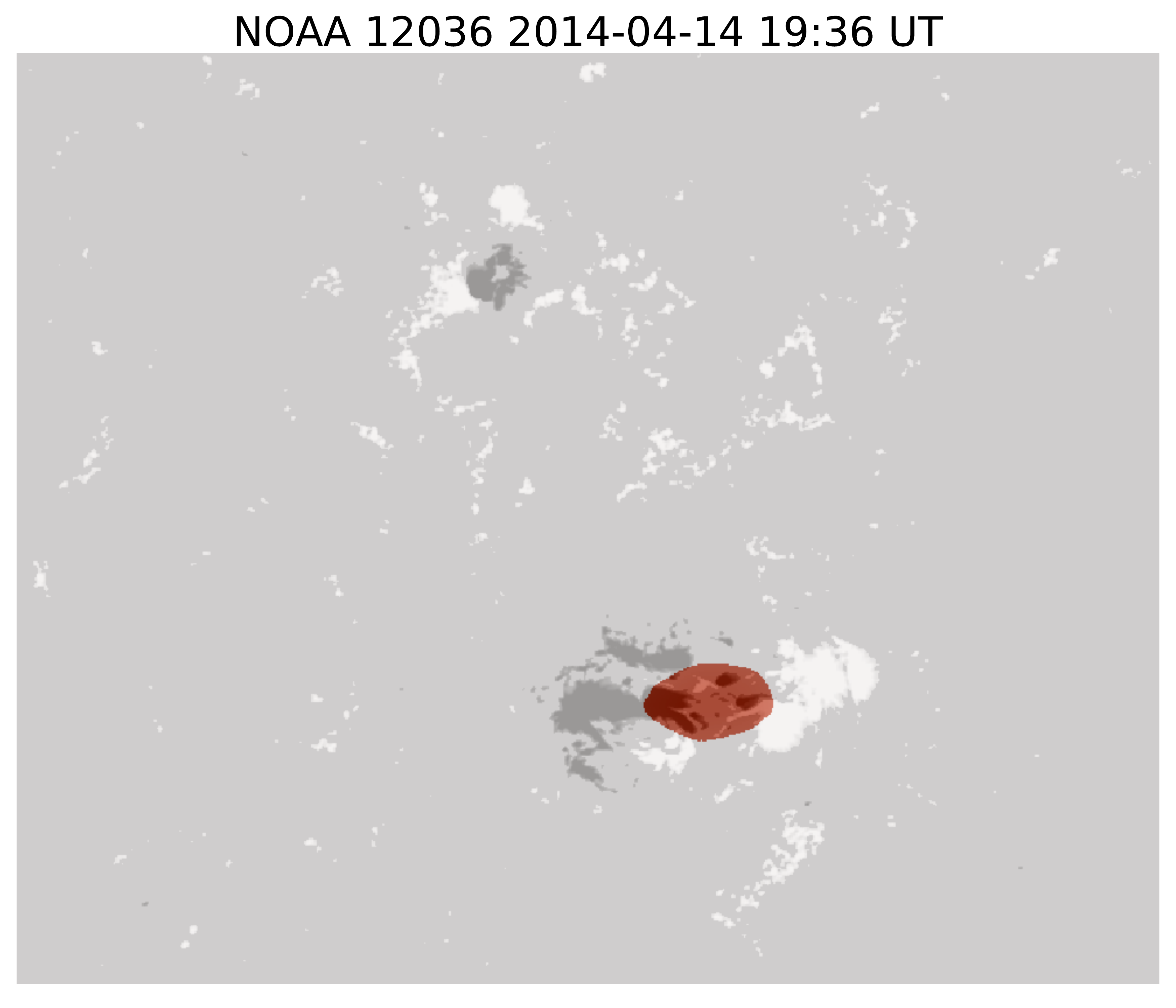}}
\caption{Visual explanations from the CNN using occlusion maps \citep{Zeiler2014} obtained by noting the change in the CNN output by systematically masking patches in the input line-of-sight magnetograms. Occlusion maps for flaring active regions (ARs) are shown. The area in {\it red} shows the part of the magnetogram where relative occlusion sensitivity is $>0.9$, i.e., regions to which the CNN output is most sensitive. For clarity in displaying the image here, magnetic fields are saturated at $+500\,{\rm G}$ ({\it white}) and $-500\,{\rm G}$ ({\it black}). In all magnetograms, the occlusion-map region highlights area between the opposite polarities \citep{Schrijver2007} irrespective of the separation distance between them.}

\label{fig:OccS}
\end{figure*}

\subsection{Probing the CNN using synthetic magnetograms \label{sec:resSyn}}
We use synthetic magnetograms to further interpret the performance of the CNN. Synthetic bipoles of circular shape with a uniform field as shown in Figure \ref{fig:SynMag} are constructed. We analyse the dependence of the CNN output on size, field strength and field configuration of synthetic bipoles.

The HMI/CEA line-of-sight magnetograms of ARs vary in sizes with a fixed spatial resolution of $0.03 \degree/{\rm pixel}$. For training, all magnetograms are resized to $256 \times 256$-pixels. Resizing yields training magnetogram images with varying spatial resolution. {\it Top} and {\it bottom} panels in Figure \ref{fig:SynResDist} show variation of the x- and y-spatial resolutions respectively of the resized images. The number of samples for flaring ARs is adjusted as per the higher penalty of misclassification levied to counter the class imbalance (see Appendix \ref{app:App1}). The mean x-resolution of the resized images is $0.11 \degree/{\rm pixel}$ and the distribution may be approximated by a wide ($\sim 0.10\degree/{\rm pixel}$) skewed Gaussian. The mean y-resolution is $0.05 \degree/{\rm pixel}$ and the distribution may be approximated by a narrower ($\sim 0.04\degree/{\rm pixel}$) Gaussian. We probe the trained CNN using synthetic bipole magnetograms with x-resolution of $0.09 \degree/{\rm pixel}$ and y-resolution of $0.05 \degree/{\rm pixel}$ ($768 \times 427$-pixels), as a representative of the training data. These results are presented in Figure \ref{fig:SynP}.

The CNN output depends on the size of the synthetic bipoles and shows low and high peaks ({\it top left}). The value of the CNN output depends on magnetic field strength. The CNN output saturates for strong magnetic fields $>3000\,{\rm G}$ ({\it top right}).  The CNN output also depends on the configuration of the synthetic bipole, favouring the configuration with $-+$ field over $+-$ field ({\it bottom left}). For a circular magnetic region with uniform $+$ or $-$ polarity, the CNN output is lower than that for the synthetic bipole with $-+$ configuration ({\it bottom right}). For a synthetic bipole of $-+$ configuration with uniform field of 2000\,G on a magnetograms with ${\rm x-res.} = 0.09\degree/{\rm pixel}$ and ${\rm y-res.} = 0.05\degree/{\rm pixel}$ ($768 \times 427$-pixels), a low peak of $\sim 0.5$ CNN output occurs at $\sim 3.5\,{\rm Mm}$ radius and a high peak of $\sim 0.8$ CNN output occurs at $\sim 12.5\,{\rm Mm}$ radius. Overall, the CNN output increases with increasing magnetic field strength and up to a certain length scale, increases with increasing size of the bipole, corroborating the dependence of CNN output on the total unsigned line-of-sight magnetic flux of ARs.

Since the resolution of magnetogram images in the training data shows a significant variation (Figure \ref{fig:SynResDist}), we explicitly study the dependence of the CNN output on resizing by comparing the CNN performance for synthetic bipole magnetograms with different x- and y-resolutions (Figure \ref{fig:SynRes}). We find that length-scales at which low and high peaks occur in the CNN outputs depend on the resolution of the resized images and in turn the magnetogram dimensions. In particular, the position of the high peak is sensitive to x-resolution of the resized magnetograms ({\it left}) and the position of the low peak is sensitive to y-resolution of the resized magnetograms ({\it center}). The length-scale of bipole at which low and high peaks occur increases with decreasing y- and x-resolutions respectively. Therefore, the CNN output for a synthetic bipole of a given radius is different depending on magnetogram dimensions, which is an artifact. We also find that the ratio of length-scales corresponding to high and low peaks is strongly correlated to the aspect ratio (x-dimension/y-dimension) of the magnetograms ({\it right}). Thus, the CNN learns to infer the resolution of the input magnetograms. The resolution of input images here is correlated with flaring activity. Low resolution implies high original dimension of the magnetogram, which corresponds to a large AR which is more likely to flare \citep{bobraflareprediction,Dhuri2019}.

We find that the asymmetry of the CNN output with respect to the polarity of magnetic fields is a consequence of the asymmetry in the number of training samples from ARs in the northern and southern hemispheres. It is known that ARs in a hemisphere have a preferred positive/negative leading polarity as per Hale's polarity law and Joy's law. By reversing the polarity of the magnetograms and retraining the CNN, we find that the CNN output curve corresponding to the $+-$ and $-+$ configuration also reverses (Figure \ref{fig:SynConfg} {\it top panel}). The number of training samples from the southern hemisphere are 25\% more than those from the northern hemispheres. We also retrained the CNN with magnetograms in the southern hemispheres modified to account for Hale's polarity law and Joy's law. To account for Hale's polarity law, we reversed the polarity of all magnetograms in the southern hemisphere. To account for Joy's law, we flipped all magnetograms in the southern hemisphere about the horizontal axis. Even after retraining in this manner, we obtained identical cross-validation TSS $\sim\,50\%$.

Thus, the trained CNN output is very sensitive to the resolution of the training images as well as the leading polarity bias. Because these factors differ significantly across different cross-validation sets, the CNN output for different cross-validation models also yields a large variation ($\sim 0.4$), as shown in the {\it bottom panel} of Figure \ref{fig:SynConfg}.

\subsection{Occlusion maps}
We generate visualisations of the CNN using tools available for qualitative interpretation of CNNs \citep{Simonyan2013,Selvaraju2017}. We use occlusion maps, which are obtained by systematically noting changes in the CNN classification label when different patches from the input magnetograms are masked \citep{Zeiler2014}. A $100 \times 100$-pixel mask is applied to generate occlusion maps from the resized magnetograms. The mask size is chosen such that the resultant change in CNN output adequately captures CNN sensitivity. For a given resized magnetogram input, the occlusion map is initialised as a $256 \times 256$ array with a uniform value of the predicted class label $Y^{\rm pred} = 1\ {\rm or}\ 0$. A $100 \times 100$-pixel region from the resized magnetogram is masked i.e. all these pixel values are set to $0$. We obtain a new predicted label $Y^{\rm mask}$. From the occlusion map, the values at the corresponding pixels are set to $Y^{\rm pred} - Y^{\rm mask}$. The process is repeated by sliding the mask by one pixel at a time such that all pixels from the input resized magnetograms are masked at least once. We also count the number of times each pixel from the resized magnetogram is masked. The net occlusion map is obtained by calculating the average value at each pixel on dividing by the counts for that pixel. The occlusion map is resized to the size of the original magnetogram using bi-cubic interpolation and normalised by the absolute maximum value.

Figure \ref{fig:OccS} shows occlusion maps for various flaring ARs. For each AR, the region with occlusion sensitivity $>0.9$, i.e., after normalisation, is shown in {\it red}. Note that the positive value of the occlusion sensitivity indicates that these regions are important for classifying regions as flaring, i.e., these regions correlate with the flaring activity. For all ARs, these regions lie between between polarities of the positive and negative magnetic fields, in the vicinity of the polarity inversion zone. This is consistent with the known nature of flux near polarity inversion line being highly correlated with flaring activity \citep{Schrijver2007,Huang2018}. Further investigation into the morphology of regions highlighted through the occlusion maps is worthwhile. However, it is required that systematic factors artificially affecting the CNN output, as described in the previous section, are eliminated before such an exercise is performed.

\begin{table*}[t]
\centering
\begin{tabular}{c>{\centering\arraybackslash}m{3.7cm}>{\centering\arraybackslash}m{3.7cm}cc}
\toprule
 & Test Dataset & Machine Inputs & Recall & TSS \\
\hline
\cite{Huang2018}& 2010 $-$ 2015 & AR magnetogram patches & 0.85 & 0.66 \\
\hline
\cite{Nishizuka_2018} & 2015 & Features extracted from magnetogram and coronal images & 0.95 & 0.80 \\
\hline
\cite{Zheng_2019} & 2010 $-$ 2018 Cross-validation & AR magnetogram patches & 0.82 $\pm$ 0.08 &  0.75 $\pm$ 0.08 \\
\hline
\cite{Li_2020} & 2010 $-$ 2018 Cross-validation & AR magnetogram patches & 0.82 $\pm$ 0.08 & 0.75 $\pm$ 0.08 \\
\hline
This work (CNN-2) & 2010 - 2015 Cross-validation & AR magnetogram patches & 0.92 $\pm$ 0.05 & 0.72 $\pm$ 0.04 \\
\hline
This work (CNN-2) & 2015 - 2018 & AR magnetogram patches & 0.86 $\pm$ 0.01 & 0.72 $\pm$ 0.01 \\
\hline
\end{tabular}
\caption{Comparison with the flare forecasting ($\geq$ M-class) performance of the trained CNN, $24~{\rm h}$ prior, with recently reported works using deep learning. The trained CNN yields a flaring class {\it recall} of $\sim 90\%$ and a {\it TSS} of $\sim 70\%$ which is comparable to the other works.}
\label{tab:performance_comp}
\end{table*}

\subsection{Comparison with other flare forecasting studies \label{sec:Results_Comparison}}
Many studies in the recent past have pursued the use of deep learning algorithms such as CNNs for flare forecasting. Although our work does not explicitly concern itself with flare forecasting, we analyse the output of the trained CNN for a given forward-looking time to infer the flare forecasting performance. In Table \ref{tab:performance_comp}, we compare {\it recall} and {\it TSS} obtained by considering magnetogram samples 24\,h before M- and X-class flares with other studies using a roughly similar approach (but different training and test datasets). We find that {\it recall} and {\it TSS} values in our study are comparable to the top-performing models. From the works reported in Table \ref{tab:performance_comp}, \cite{Nishizuka_2018} yield the best classification {\it TSS} of $\sim 80\%$. They follow a slightly different approach than ours, using features extracted from AR magnetograms (as well as coronal images), rather than directly training the neural network on AR magnetograms. Other works mentioned in Table \ref{tab:performance_comp} take a similar approach and also yield broadly similar results. Therefore, our findings of the operation of the trained CNN and artifacts that arise as a result of resizing the images are likely applicable to these studies as well.

\section{Summary \label{sec:summary}}
We have successfully trained CNNs to distinguish between SDO/HMI line-of-sight magnetograms of flaring and nonflaring ARs. We trained two CNNs --- a baseline model with a simple architecture, CNN-1 (Figure \ref{fig:cnn1}) and a complex model with inception modules, CNN-2 (Figure \ref{fig:cnn2}). We find that CNN-2 performs significantly better than CNN-1, yielding 10-fold cross-validation {\it TSS} of $\sim 50\%$. We also calculated instantaneous  {\it recall} of flaring ARs between $\pm 72$ hours of M- and X-class flares. We find that the {\it recall} for flaring ARs peaks at $\sim 80\%$, which is consistent with reported results for forecasting flares using CNNs trained on line-of-sight magnetograms \citep{Huang2018}. The peak {\it recall} value obtained using an SVM trained on vector-magnetic-field features averaged over ARs is $\sim 10\%$ higher \citep{Dhuri2019}. Also, an SVM trained with AR features that may be reliably inferred from line-of-sight magnetograms gives results similar to the CNN. Thus, the trained CNN is mainly looking at the global (extensive) features of the ARs, such as AR area and total magnetic flux, for the classification. The instantaneous {\it recall} may be interpreted as a correlation between the line-of-sight magnetic field and flaring activity.  Consistent with \cite{Dhuri2019}, the instantaneous {\it recall} for the CNN, determined primarily by the extensive features, stays high, $>50\%$, for days before and after flares. Analysed for forecasting $\geq$ M-class flares 24\,h prior, the CNN yields a {\it recall} of $\sim$ 90\% and a {\it TSS} of $\sim$ 70\%, which is also comparable to the reported results using CNNs.

We performed a statistical analysis of total unsigned line-of-sight flux of flaring and nonflaring AR magnetograms binned by CNN output. We find that the average value of total unsigned line-of-sight flux increases as CNN output increases. This suggests that the total unsigned line-of-sight flux --- an extensive AR feature --- strongly dictates the CNN performance. Using synthetic magnetograms, we find that the CNN output shows low and high peaks at two different length scales of synthetic bipoles. We show that synthetic bipole length scales corresponding to the low and high peaks are a characteristic of y- and x-resolutions respectively of the resized input magnetograms. We also obtained visualisations from the CNN using occlusion maps \citep{Zeiler2014}. The occlusion maps show the region in the magnetograms to which the CNN output is most sensitive. We find that this region lies between opposite polarities of ARs, irrespective of how far the polarities are spatially separated. This is consistent with earlier studies that show the flux near the polarity inversion line as being strongly correlated with flaring activity \citep{Schrijver2007}.

A detailed analysis of the morphology of regions highlighted by occlusion maps may shed light on the role of the polarity inversion zone in triggering flares. We find, however, using synthetic magnetograms, that the CNN output depends on spurious factors such as the magnetogram dimensions (for same-sized magnetic regions). This is a direct consequence of the resizing operation used to prepare AR magnetograms in identical sizes as an input to the CNN. We also find that the CNN output is asymmetric with respect to the polarity of the magnetic field which is due to the asymmetry in the number of training samples of ARs from the northern and southern hemisphere.
A CNN design that eliminates these systematic effects has the potential to reveal new morphological characteristics of flaring ARs. Including additional inputs to the CNN, such as coronal and chromospheric imagery, will also improve the CNN characterisation of local AR features for understanding and forecasting flares. Insights from this study about the operation of the CNN for AR magnetograms will be useful for future applications that include more comprehensive AR information for reliable flare forecasting.

\acknowledgments
D.B.D and S.M.H designed the research. S.B, R.A., and D.B.D. performed the data analysis. D.B.D and S.M.H. contributed to the interpretation of the results. D.B.D. wrote the manuscript with contributions from all authors.

S.M.H acknowledges funding support from the Max-Planck partner group program. HMI data used here are courtesy of NASA/SDO and the HMI science team. We thank the anonymous reviewer for their comments and suggestions which helped improve the analysis and presentation of the paper.


\appendix
\section{Details of the Neural Networks \label{app:App1}}
Neural networks consist of layers of neurons. A neuron in each layer processes N-dimensional inputs $\textbf{x}$ according to the following operation and produces an output $y$
\begin{equation} \label{eq:neuronAct}
    y = f\left(\sum_{j=1}^{N} w_j x_j + b\right).
\end{equation}
Here the N-dimensional vector $\textbf{w}$ contains weights of the neuron and $b$ is the bias of the neuron. The input to the first layer of neurons is the data. Outputs from a layer of neurons serve as inputs to the next layer. The final layer of neurons provides the output of the neural network. As explained in Section \ref{sec:method}, a convolutional neural network comprises convolutional filters with ${\rm M} \times {\rm M}$ neurons that slides over the input image data. Function $f$ in Eq. \ref{eq:neuronAct} is the neuron {\it activation function}. Since the magnetograms contains pixels with both positive and negative magnetic field, we use {\it leaky ReLU} (rectified linear unit) activation function \citep{maas2013rectifier}. For CNN-2, we find that the identity activation function,
\begin{equation} \label{eq:identityAct}
    y = \sum_{j=1}^{N} w_j x_j + b,
\end{equation}
also yields similar results.

\begin{figure}[t]
\centering
  \includegraphics[width=0.60\textwidth,trim={0cm 0cm 0cm 0cm},clip]{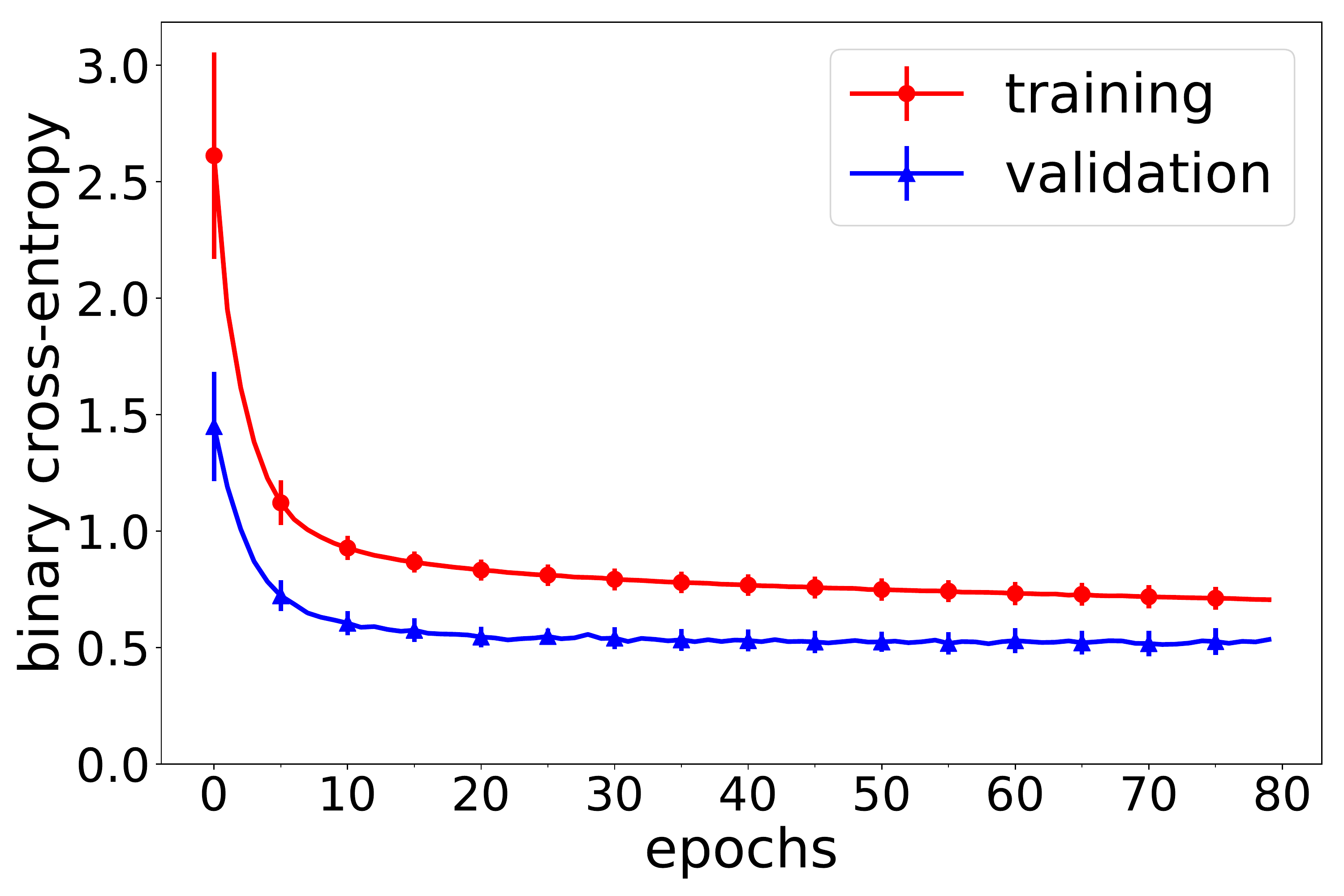}
\caption{Variation of mean cross-validation training and validation loss with the progression of training epochs for the CNN-2. The variation is similar for both training and validation indicating no/minimal overfitting. $1\sigma$ error bars are shown. \label{fig:appTraining}}
\end{figure}

The final layers of CNN-1 and CNN-2 produce output $Y^{\rm pred}$ which is a number between $0$ and $1$. We compute a loss function  $L(Y,Y^{\rm pred})$ to determine the error in the prediction. Weights and biases of all neurons in the CNN are determined during training using minibatch stochastic gradient descent to minimize the loss $L(Y,Y^{\rm pred})$ (see Figure \ref{fig:appTraining}). Stochastic gradient descent is performed using a minibatch of samples $n$ during each training step. We use the binary cross-entropy loss function given by \citep{hastie01statisticallearning}. 
\begin{equation} \label{eq:CELossFunction} 
    L_{\rm CE}(Y,Y^{\rm pred}) = -\frac{1}{n} \sum_{i=1}^{n} \left[ CY_{i} \log \left(Y^{\rm pred}_i\right) + \left(1-Y_i\right) \log \left(1-Y^{\rm pred}_i\right) \right]. 
\end{equation}
Here, $C$ is the additional penalty for the misclassification of positive, i.e., flaring-class magnetograms labelled as 1. We set the value of $C=4.5$ which is approximately equal to the class-imbalance ratio, i.e., the ratio of nonflaring to flaring samples in the data (see Table \ref{tab:dataset}). For the baseline model (CNN-1), we find the best performance (noted in Table \ref{tab:cnn_comp}) with a custom loss function 
\begin{equation} \label{eq:CULossFunction}
    L_{\rm custom}(Y,Y^{\rm pred}) = \frac{1}{n_{\rm fl}} \sum_{i=1}^{n_{\rm fl}} \lambda (Y_i^{\rm fl} - Y_i^{\rm fl,pred})^2 + \frac{1}{n_{\rm nfl}} \sum_{i=1}^{n_{\rm nfl}} (Y_i^{\rm nfl} - Y_i^{\rm nfl,pred})^2,
\end{equation}
where 
\begin{equation} \label{eq:multFact}
    \lambda = C_{\rm batch} \exp({\rm HSS_{\rm batch}}) \exp({\rm TSS_{\rm batch}}).
\end{equation}
In the above equation, $C_{\rm batch}$ is the class-imbalance ratio for the minibatch used, TSS is the true-skill-statistic (see Section \ref{sec:method}) for the minibatch, and HSS is Heidke Skill Score for the minibatch given by \citep{bobraflareprediction}
\begin{equation} \label{eq:HSS}
\text{HSS} = \frac{2(TP.TN - FP.FN)}{(TP+FN)(FN+TN)+(TP+FP)(FP+TN)}.
\end{equation}
For the present classification problem, we threshold $Y^{\rm pred}$ to obtain the predicted class labels as
\begin{equation} \label{eq:outputTH}
    Y^{\rm pred} =  \left\{
    \begin{array}{ll}
    1 & \text{ if } Y^{\rm pred} \ge  0.5, \\
    0 & \text{ else} .
    \end{array}
\right.
\end{equation}

We use learning rate $lr=5 \times 10^{-6}$ and a minibatch size of $n=64$ for the stochastic gradient descent. We also use batch normalisation to pre-process inputs to each convolution layer \citep{ioffe2015batch}.
\end{document}